\documentclass{article}
\usepackage{authblk}
\usepackage[utf8]{inputenc}
\usepackage[left=2cm,right=2cm,top=1.5cm,bottom=1.5cm]{geometry}
\usepackage{amsmath}
\usepackage{amssymb}
\usepackage{graphicx}
\usepackage{subfig}
\usepackage{mathtools}
\usepackage{mathpazo}
\usepackage[mathpazo]{flexisym}
\usepackage{breqn}
\usepackage{comment}
\usepackage{verbatim}
\usepackage[dvipsnames]{xcolor}
\usepackage{placeins}
\usepackage{hyperref}
\usepackage{soul}
\usepackage[normalem]{ulem} 
\hypersetup{
    colorlinks=true,
    linkcolor=blue,
    filecolor=magenta,      
    urlcolor=cyan,
    citecolor=teal
    }
\usepackage{comment}
\graphicspath{{figs/}}
\usepackage[style=numeric-comp,style=phys,url=false]{biblatex}
\DeclareMathOperator{\sech}{sech}

\title{Collective coordinate descriptions of a kink in a driven-damped $\phi^4$ model}
\author[1]{Jacek Gatlik}
\author[2]{Tomasz Dobrowolski}
\author[3]{Jean-Guy Caputo}
\author[4]{Panayotis G. Kevrekidis}
\affil[1]{\textit{AGH University of Krakow, Faculty of Physics and Applied Computer Science, 30-059 Krakow, Poland}}
\affil[2]{\textit{Department of Physics and Applied Mathematics, University of the National Education Commission, Podchor\c{a}\.zych  2, 30-084 Krakow, Poland}}
\affil[3]{\textit{Laboratoire de Math\'{e}matiques, INSA de Rouen, Avenue de l'Universit\'{e}, 76801 Saint-Etienne du Rouvray, France}}
\affil[4]{\textit{Department of Mathematics and Statistics, University of Massachusetts, Amherst, Massachusetts 01003-4515, USA}}
\date{\today}

\begin{document}
\maketitle

\begin{abstract}
Extending a recent effective theory formulation for the dynamics of kinks in
the sine-Gordon model \cite{Dobrowolski2025}, we propose an
analogous effective description of $\phi^4$ kinks.
Three different reduced models based on the kink position, width and
internal mode amplitude are introduced and compared systematically
with the numerical solution of the equation with space- and time-dependent
perturbations.
In all cases considered, the model based on the kink position and width agrees the best
with the full numerical solution. As long as the external driving
frequency of the perturbation remains moderate, it captures with remarkable
accuracy the intricate dynamical processes taking place in the system.
\end{abstract} \hspace{10pt}

\begingroup
\footnotesize
\setlength{\parskip}{0pt}
\tableofcontents
\endgroup

\section{Introduction}

Many physical phenomena are elegantly modeled by scalar field theories that support topological solitons such as kinks (in 1+1 dimensions) or domain walls (in 3+1 dimensions) \cite{Caputo2024,Ma2025}. Among these, the $\phi^4$ model stands as one of the simplest yet most widely applicable frameworks. Its nonlinear double-well potential generates kink solutions that emerge across multiple disciplines~\cite{p4book}.

In condensed matter physics, $\phi^4$ field theory serves as a paradigmatic framework for describing domain walls and other nonlinear excitations in ferroic systems \cite{Chaikin1995,Vachaspati2023}. Classic treatments emphasize how the quartic potential with spontaneous symmetry breaking captures the essential energetics, stability, and profile of domain walls, linking universal features of soliton physics with material-specific realizations \cite{Rajaraman1982}. For ferroelectrics, $\phi^4$-based continuum models have been extensively used to rationalize the structure, width, and dynamics of walls, complementing both microscopic approaches and phenomenological Ginzburg–Landau–Devonshire theories \cite{Catalan2012,Meier2021}. Analogous $\phi^4$ descriptions extend to ferroelastic transitions and mechanical metamaterials, where quartic-order energy landscapes underlie pattern formation and mechanically tunable domain-wall–like states \cite{Salje2012,Bertoldi2017}. Even in photonics and optoelectronics, the interaction of light with ferroic domain walls has been analyzed through the lens of $\phi^4$-type soliton dynamics, highlighting the broad applicability of the model across disciplines \cite{Seidel2016,Nataf2020}.

In biological and ecological systems, $\phi^4$-inspired frameworks can describe interfaces, pattern formation, and propagation between states (e.g., active/inactive regions, healthy/diseased tissue), especially when modeled via reaction–diffusion equations~\cite{Xin2000Fronts}. Although explicit $\phi^4$ usage in such contexts remains sparse, the conceptual parallels-front propagation, bistability, and interface motion-are well-established and suggest fertile ground for cross-disciplinary modelling \cite{Saarloos2003}.

Beyond their classical applications, $\phi^4$-type models play an important role in statistical physics and the study of phase transitions. In particular, mean-field $\phi^4$ systems allow for analytic evaluation of configurational entropy and provide insight into nonanalytic behavior, thereby shedding light on the nature of continuous transitions without the need to invoke topology changes in configuration space \cite{Hahn2005}.

Non-autonomous $\phi^4$ models-those with explicit time- or space-dependent parameters-represent an emerging and particularly exciting direction. For instance, models under external fields reveal rich behavior including periodic domain-wall arrays, soliton interactions, and field-induced shifts, as shown via exact solutions in coupled $\phi^4$ models \cite{Braghin1998,Destri2000}. While direct applications of non-autonomous  $\phi^4$ in biology or condensed matter are just beginning to appear, advances in non-autonomous soliton theory-especially within nonlinear Schrödinger frameworks-provide powerful mathematical tools to construct time-dependent soliton solutions. This suggests that applying similar techniques to $\phi^4$  systems with modulated or driven parameters may soon yield valuable descriptions of interfaces in dynamic or forced media.
It is that motivation that drives the considerations presented herein, where
the evolution of a $\phi^4$ kink under such spatio-temporal perturbations
is examined.

The paper is organized as follows. In \hyperref[sec2]{Sec. 2}, we introduce the model we investigate with both space and time perturbations. \hyperref[sec3]{Section 3} is devoted to the construction of three effective models, each based on a different ansatz. These three models are compared to the field solution in both the autonomous and non-autonomous cases (without dissipation). In \hyperref[sec4]{Sec. 4}, an effective model incorporating dissipation and drive is developed and subsequently tested against the field-theoretic results. Finally, \hyperref[sec5]{Sec. 5} contains a discussion of the main findings.
\section{The driven damped $\phi^4$}
\label{sec2}
We consider the $\phi^4$ model defined as
\begin{equation}
\label{phi4+}
\partial_t^2\phi + \eta \partial_t \phi - \partial_x(\mathcal{F}(x)\partial_x\phi) + \lambda(t,x) \, \phi (\phi^2 -1)  = -\Gamma,
\end{equation}
where the parameter $\eta$ accounts for dissipative effects within the system. The coupling $\lambda$ is taken to be a function of both time and space, reflecting the influence of external modulation. In this work, we consider specific forms for the functions ${\mathcal F}(x)$ and $\lambda(t,x)$.
Motivated by the fact that elemental Fourier building blocks can
be used to represent more complex spatial and temporal dependences, 
we choose ${\mathcal F}(x)$  to be a spatially periodic function, while 
$\lambda(t,x)$ is selected as a traveling wave characterized by a frequency $\omega$ and a wave number $k$
\begin{equation}
\label{F_function}
\mathcal{F}(x) = 1+\varepsilon_1\sin(\frac{\pi}{12}x) \mathrm{ ~~~~ and ~~~~ } \lambda(t,x) = 1 + \varepsilon_2\sin(kx-\omega t) .
\end{equation}
In this work we employ two types of initial conditions, chosen so as to remain consistent with the adopted ans{\"a}tze. The first set follows the natural scalings present in the modified ansatz and is given by the system of equations:
\begin{equation}\label{phi_wp1}
\phi(0,x)=  \tanh \left[  \sqrt{\frac{\lambda(t=0,x_0)}{{2 \mathcal{F}(x_0)}}} \gamma_0 \, (x-x_0)  \right],
\end{equation}
\begin{equation}\label{phi_wp2}
\partial_t \phi(0,x)= -v \sqrt{\frac{\lambda(t=0,x_0)}{{2 \mathcal{F}(x_0)}}} \gamma_0 \,\sech^2 \left[   \sqrt{\frac{\lambda(t=0,x_0)}{{2 \mathcal{F}(x_0)}}} \gamma_0 \, (x-x_0)\right],
\end{equation}
where $\gamma_0=1/\sqrt{1-v^2}$. Here, the parameter $v$ denotes the initial velocity of the kink. This particular choice of initial conditions ensures consistency between the initial field configuration and the background structure shaped by the system's inhomogeneities, as well as the ansatz employed for constructing a meaningful effective description.   
The second set of initial conditions is adapted to the traditional ansatz
\begin{equation}\label{phi_wp-x}
\phi(0,x)=  \tanh \left[  \frac{1}{\sqrt{2}} \gamma_0 \, (x-x_0)
\right], \,\,\,\,
\partial_t \phi(0,x)= -v \frac{1}{\sqrt{2}} \gamma_0 \, \mathrm{sech}^2 \left[   \frac{1}{\sqrt{2}} \gamma_0 \, (x-x_0)\right],
\end{equation}
here $\mathcal{F}$ and $\lambda$ are different from $1$ in our 
inhomogeneous domain, while at the domain 
boundaries Dirichlet boundary conditions
consonant with the kink solution are imposed.

\section{Three effective models in the non-dissipative case}
\label{sec3}
In this section, we introduce three  effective models to describe the dynamics of a kink in our $\phi^4$ model. One model is based on the variable $x_0$, the position of the kink, and the variable $\gamma$, the inverse of the kink width. The next two models contain the variable $x_0$ and a dynamic variable $b$ determining the amplitude of vibrations of the discrete mode. 

\subsection{Model 1 - position and width $(x_0,\gamma)$}
In the absence of both dissipation $\eta$ and external forcing $\Gamma$, equation \eqref{phi4+} simplifies to:
\begin{equation}
\label{phi4}
\partial_t^2\phi  - \partial_x(\mathcal{F}(x)\partial_x\phi) + \lambda(t,x) \, \phi (\phi^2 -1)  = 0,
\end{equation}
The corresponding equation of motion is derived from the following Lagrangian:
\begin{equation}\label{L}
L= \int_{-\infty}^{+\infty} dx {\cal L}(\phi)
=\int_{-\infty}^{+\infty} dx  \left[ \frac{1}{2}~ (\partial_t
\phi)^2 - \frac{1}{2} ~{\cal F}(x) (\partial_x \phi)^2 - V(\phi) \right],
\end{equation}
where the potential $V$ is
\begin{equation}\label{V}
V(\phi) = \frac{1}{4} \lambda(t,x) \left(\phi^2 - 1 \right)^2 .
\end{equation}
We introduce  model 1 by rewriting the field in terms of a new field variable $\xi(t,x)$, defined as:
\begin{equation}\label{kink-xi}
\phi_K(t,x) =  \tanh \xi(t,x).
\end{equation}
The functional form of $\xi(t,x)$ is to be derived by requiring that it satisfies its corresponding equation of motion. Plugging $\phi = \phi_K$ into  \eqref{L} and computing the integrals, we obtain the Lagrangian:
\begin{equation}
\begin{gathered}
\label{Lxi}
L(\xi, \partial_{t} \xi,t) = \int_{-\infty}^{+\infty} dx  \,\, {\rm sech}^4 \xi \left(
\frac{1}{2}\,   (\partial_{t} \xi)^2 
-   \frac{1}{2}\, {\cal
F}(x)  (\partial_{x} \xi)^2 
 - \frac{1}{4}~\lambda(t,x) 
   \right).
\end{gathered}
\end{equation}
Up to this point, the transformation given in \eqref{kink-xi} has left the field dynamics of the model entirely unaffected.
To proceed, following the $x,t$ scalings, we impose the ansatz
\begin{equation}\label{xi}
 \xi= \sqrt{ \frac{\lambda(t,x_0(t))}{2{\cal F}(x_0(t))}} \,\, \gamma(t) \, (x-x_0(t)),
\end{equation}
thereby restricting our analysis to the sector that includes the kink solution. From this point forward, the system’s dynamics is fully characterized by just two time-dependent quantities: the kink’s center $x_0(t)$ (defined as the point where the scalar field vanishes) and the quantity related to the inverse width of the kink $\gamma(t)$. {The adoption of this ansatz is dictated by the analogy to the previously studied sine-Gordon model \cite{Dobrowolski2025b}. In that model, a similar ansatz allowed for an excellent description of the kink trajectory and changes in its width in a non-autonomous system rich in spatial inhomogeneities. It is important to note that one of the key objectives of this work is to test the generality of this ansatz, i.e., how effectively it captures kink dynamics across models with spatial inhomogeneities.} Integrating over space then yields an effective Lagrangian of the form
\begin{equation}\label{Leff}
L_{eff} (x_0, \gamma, \Dot{x}_0, \Dot{\gamma},t)= \frac{1}{2}\, M \Dot{x}_0^2 +\frac{1}{2}\, m
\Dot{\gamma}^2 + \kappa \Dot{x}_0 \Dot{\gamma} + \alpha \Dot{x}_0
+ \beta \Dot{\gamma} - V.
\end{equation}
The coefficients featured in the Lagrangian are detailed in \hyperref[AppendixA]{Appendix A}. They depend on the variables $x_0$,  $\gamma$  and inherit an explicit time dependence through the time-varying coupling "constant"  $\lambda$. {At this point, it is worth noting that the effective Lagrangian of the $\phi^4$ model we obtained is almost identical to a similar Lagrangian obtained in the case of the sine-Gordon model \cite{Dobrowolski2025b}. The only fundamental difference is the functional dependence of the potential and the coefficients of the effective model on the dynamic variables $(x_0,\gamma)$.} Given this Lagrangian, one can derive the Euler–Lagrange equations governing the system’s dynamics
\begin{equation}
\begin{gathered}
\label{2dof_ansatz}
    M\Ddot{x}_0+\kappa \Ddot{\gamma}+ (\partial_{\gamma} M) \Dot{x}_0 \Dot{\gamma}+\frac{1}{2}(\partial_{x_0}M)\Dot{x}_0^2+\left(\partial_{\gamma} \kappa-\frac{1}{2}\partial_{x_0}m \right)\Dot{\gamma}^2+ \\
    (\partial_t M) \Dot{x}_0 +\left(\partial_{\gamma} \alpha + \partial_t \kappa -\partial_{x_0} \beta \right) \Dot{\gamma} + \partial_t \alpha +  \partial_{x_0}V=0,\\
    m\Ddot{\gamma} + \kappa \Ddot{x}_0+ (\partial_{x_0} m) \Dot{x}_0 \Dot{\gamma}+\frac{1}{2}(\partial_{\gamma}m)\Dot{\gamma}^2
    + \left( \partial_{x_0} \kappa
    -\frac{1}{2}\partial_{\gamma}M \right)\Dot{x}_0^2 +\\
(\partial_t m) \Dot{\gamma} + \left(\partial_{x_0} \beta + \partial_t \kappa - \partial_{\gamma} \alpha \right) \Dot{x}_0  +\partial_t \beta  +   \partial_{\gamma}V=0.
\end{gathered}
\end{equation}
Naturally, the partial time derivatives appearing in the equations account solely for the time dependence introduced explicitly via $\lambda$.

\FloatBarrier
\subsection{Model 2 position and internal mode $(x_0,b)$ }
We build {\rm model 2} using an ansatz that contains two dynamic variables: the position of the kink $x_0(t)$, and the amplitude $b$ of the discrete vibrating mode part of the spectrum of linear excitations of the unperturbed $\phi^4$ model:
\begin{equation}\label{kink-xi2++}
\phi_K(t,x) =  \tanh \xi(t,x) + b(t) \tanh \xi(t,x) \sech \xi(t,x),
\end{equation}
where
\begin{equation}\label{xi2++}
 \xi=  \frac{1}{\sqrt{2}} \,\, (x-x_0(t)).
\end{equation}
It is worth noting that an ansatz analogous to Eq. \eqref{kink-xi2++} is also employed in the context of describing kink–antikink interactions. In this setting, however, it leads to serious difficulties. In particular, the standard Sugiyama ansatz gives rise to the so-called null-vector problem, i.e., a degeneracy of the collective coordinates and a singularity of the induced metric in the vicinity of vanishing kink–antikink separation, as clearly analyzed in Ref. \cite{Takyi2016}. This issue was later reinterpreted as an artifact of an inappropriate choice of coordinates and resolved by a suitable reparametrization of the moduli space \cite{Manton2021}. The effective Lagrangian obtained on the basis of the above ansatz takes the form
\begin{equation}\label{Leff++}
L^b_{eff}(x_0, b, \Dot{x}_0, \Dot{b},t) = \frac{1}{2}\, M_b \Dot{x}_0^2 +\frac{1}{2}\, m_b
\Dot{b}^2   - V_b.
\end{equation}
The mass terms present in this Lagrangian are defined below. Note that the mass $M_b$ associated with the position variable is a function of the variable $b$
\begin{equation}
\label{Mb++}
    M_b = \int_{-\infty}^{+\infty}dx \frac{1}{2}~{\mathcal{K}^2} = \frac{1}{\sqrt{2}} \left( \frac{4}{3} + \frac{14}{15} \, b^2 + \frac{\pi}{2} b \right) ,
\end{equation}
where $\mathcal{K}$ is defined as follows
\begin{equation}
    \mathcal{K} = \sech^2 (\xi) + b \sech(\xi) \left( -1 + 2 \sech^2 (\xi) \right) .
\end{equation}
On the other hand, the mass associated with the variable $b$ is constant 
\begin{equation}
\label{mb++}
    m_b = \int_{-\infty}^{+\infty}dx ~ \tanh^2(\xi) \sech^2(\xi) = \frac{2 \sqrt{2}}{3} .
\end{equation}
The effective potential contains both functions, i.e., $\mathcal{F}(x)$ and $\lambda(t,x)$, causing deviations of the field model Eq.~(\ref{phi4}) from the original $\phi^4$ model. These functions are responsible for the explicit breaking of translational invariance and the non-autonomy of the model  
\begin{equation}
\label{Vb++}
    V_b = \int_{-\infty}^{+\infty}dx \,   \,  \left[  \frac{1}{2} \,  \mathcal{F}(x)  \frac{1}{2} \mathcal{K}^2\, + \frac{1}{4} \lambda(t,x) \left( \tanh^2(\xi) \left( 1 + b \sech(\xi)\right)^2-1 \right)^2\right].
\end{equation}
The equations of motion of the effective model constitute a system of nonlinear and coupled second-order ordinary differential equations of the form
\begin{equation}
\begin{gathered}
\label{2dof_ansatz++}
    M_b\Ddot{x}_0+(\partial_{b} M_b) \Dot{x}_0 \Dot{b}+\partial_{x_0}V_b=0,\\
    m_b\Ddot{b} -
    \frac{1}{2} \left( \partial_{b}M_b \right)\Dot{x}_0^2+\partial_{b}V_b=0.
\end{gathered}
\end{equation}
In the following we  compare the field to the predictions of model 2. For this we employ the initial conditions specified by the system of equations \eqref{phi_wp-x} because they are the only ones that can be consistently realized within the ansatz adopted in this part of the study.

\subsection{Model 3 - position and internal mode $(x_0,b)$}
In anticipation of the forthcoming numerical results, and bearing in mind the deviations exhibited by the second model with respect to the predictions derived from the field-theoretic approach Eq. \eqref{phi4}, we also investigated a modified version of model 2 that explicitly incorporates the presence of inhomogeneities in the system.
This model 3 with two degrees of freedom is based on the following ansatz:
\begin{equation}\label{kink-xi2}
\phi_K(t,x) =  \tanh \xi(t,x) + b(t) \tanh \xi(t,x) \sech \xi(t,x) .
\end{equation}
As was the case with the previous model, the second term is associated with a bound state mode in the linear spectrum of the $\phi^4$ model. The amplitude $b(t)$ accompanying this term serves as the second dynamical variable in the present framework. The position of the kink $x_0(t)$, serving as the first dynamical variable  is defined through the following expression:
\begin{equation}\label{xi2+}
 \xi= \sqrt{ \frac{\lambda(t,x_0(t))}{2 {\cal
F}(x_0(t))}} \,\, (x-x_0(t)).
\end{equation}
It is worth noting that, already at the level of the ansatz, we incorporate functions that introduce spatial or temporal heterogeneity into the system.
{This modified form of the variable $\xi$ is motivated by analogy with the model proposed in article \cite{Dobrowolski2025b}.}
Integration with respect to the spatial variable leads to the effective Lagrangian:
\begin{equation}\label{Leff+}
L^b_{eff} (x_0, b, \Dot{x}_0, \Dot{b},t)= \frac{1}{2}\, M_b \Dot{x}_0^2 +\frac{1}{2}\, m_b
\Dot{b}^2 + \kappa_b \Dot{x}_0 \Dot{b} + \alpha_b \Dot{x}_0
+ \beta_b \Dot{b} - V_b.
\end{equation}
Although the general form of the Lagrangian is similar to that in the first model, the functional dependence of the coefficients on the dynamic variables $x_0(t)$ and $b(t)$ is significantly different and is detailed in \hyperref[AppendixB]{Appendix B}. 
Obviously, the Lagrangian involves a set of coefficients which depend on the dynamical variables. Owing to the explicit time dependence of the "coupling" coefficient $\lambda$, this temporal variation is inherited by all the coefficients in question. 
Once the Lagrangian structure is established, the Euler–Lagrange equations governing the system follow in a straightforward manner
\begin{equation}
\begin{gathered}
\label{2dof_ansatz--} M_b\Ddot{x}_0+\kappa_b \Ddot{b}+ (\partial_{b} M_b) \Dot{x}_0 \Dot{b}+\frac{1}{2}(\partial_{x_0}M_b)\Dot{x}_0^2+\left(\partial_{b} \kappa_b-\frac{1}{2}\partial_{x_0}m_b \right)\Dot{b}^2+ \\
    (\partial_t M_b) \Dot{x}_0 +\left( \partial_t \kappa_b + \partial_b\alpha_b -\partial_{x_0} \beta_b \right) \Dot{b} + \partial_t\alpha_b +  \partial_{x_0}V_b=0,\\
    m_b\Ddot{b} + \kappa_b \Ddot{x}_0+ (\partial_{x_0} m_b) \Dot{x}_0 \Dot{b}  +\frac{1}{2}(\partial_b m_b)\Dot{b}^2
    + \left( \partial_{x_0} \kappa_b - \frac{1}{2}\partial_b M_b\right)\Dot{x}_0^2 +\\
(\partial_t m_b) \Dot{b} + \left( \partial_t \kappa_b + \partial_{x_0}\beta_b - \partial_{b} \alpha_b \right) \Dot{x}_0  +\partial_t\beta_b +   \partial_{b}V_b=0.
\end{gathered}
\end{equation}
{We shall compare the predictions derived from this set of equations with those obtained from the model discussed in the previous section. Let us note that, similarly to the first model, this model can be studied using the system \eqref{phi_wp1}, \eqref{phi_wp2} of initial conditions. These conditions take into account the inhomogeneity of the system, in the same way as the adopted ansatz. On the other hand, since the initial velocities of the kink are zero or nearly zero, in the third model at the beginning of the evolution $\gamma$ is equal or nearly equal to one (the only exception being the example shown in Fig. \ref{nfig_03} where $v=0.0001$). This allows us to directly compare the first and third models.}

\section{Numerical validation of the effective models in non-dissipative case}
\label{sec4}
In this section, we compare the  effective models 1, 2, and 3 with the field solution of equation \eqref{phi4}. Since at zero initial velocities we can set initial conditions in the effective models consistent with the conditions described by equations \eqref{phi_wp1} and \eqref{phi_wp2}, we always compare the accuracy of these models in terms of the evolution of the variable $x_0$ in a single figure.
 The situation is slightly different for model 2, whose ansatz is consistent with the initial conditions \eqref{phi_wp-x}. In this case, we compare changes in the position of the kink only with the field solution of the equation \eqref{phi4}. However, all required parameters are selected identically in both cases.
\subsection{Numerical results for spatial modulations of ${\mathcal F}$ and $\lambda$}
First, we present typical results obtained in  model \eqref{phi4}, or equivalently in  model \eqref{phi4+}, without dissipation ($\eta=0$) and external forcing ($\Gamma=0$).

\subsubsection{Spatial modulation of  ${\mathcal F}$}
\begin{figure}[h!]
    \centering
    \subfloat{{\includegraphics[height=4.2cm]{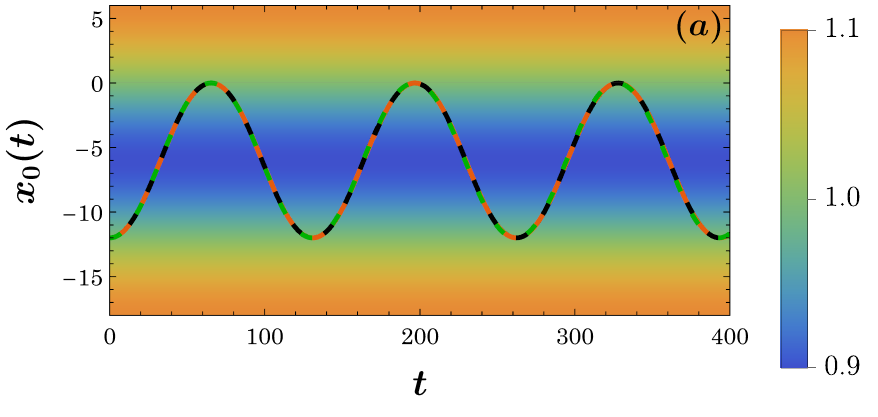}}}
    \quad
    \subfloat{{\includegraphics[height=4.5cm]{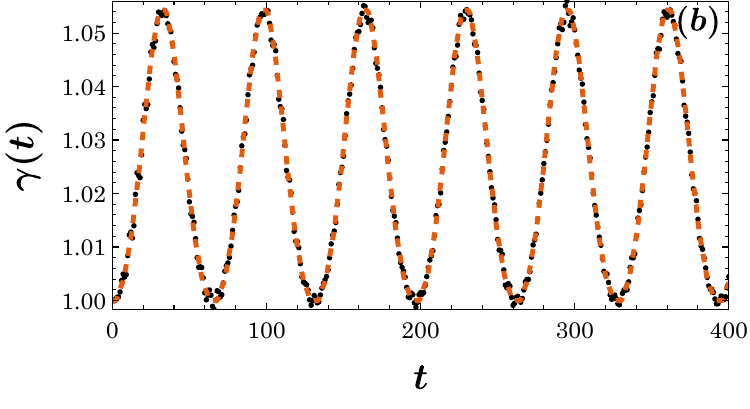}}}
    \quad
    \subfloat{{\includegraphics[height=4.5cm]{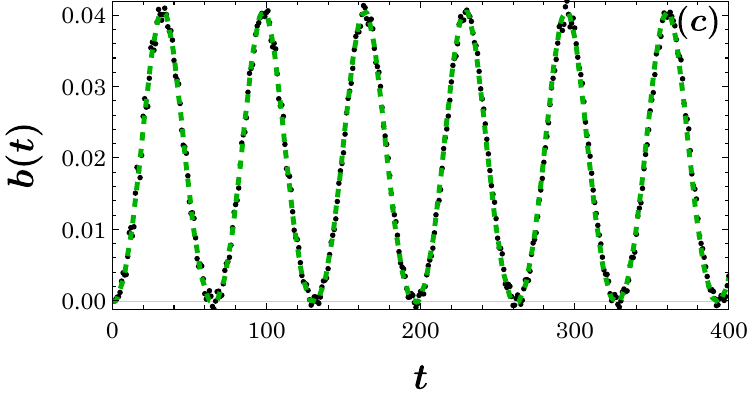}}}
    \caption{The trajectory of a kink that initially rests ($v=0$) at point $x_0=-12$. In this simulation, $\varepsilon_2=0$, which means that $\lambda=1$. The heterogeneity in the system is described only by the function $\mathcal{F}$ through the value of the parameter $\varepsilon_1=0.1$. (a) Comparison of the trajectory obtained from the field equation \eqref{phi4} (black solid line), the first effective model (red dashed line), and the third approximate model (green dashed line). The color bar on the right side of 
    panel (a) refers to the values of the function $\mathcal{F}$. (b) Comparison of the variable $\gamma$ obtained from the field equation (black dotted line) and obtained from the first effective model (red dotted line). (c) Variable $b$ obtained from the field equation (black dotted line) and the third approximate model (green dotted line).}
    \label{nfig_01}
\end{figure}

Let us first consider the case where $\lambda=1$, i.e., the parameter preceding the potential does not depend on time or on the spatial variable. The function ${\mathcal F}$ is a periodic function. Figure \ref{nfig_01} depicts the oscillations of the kink between two adjacent maxima of the function ${\mathcal F}$. The colorbar on the side of the figure illustrates the 
range of values of the function ${\mathcal F}$. In fact, the oscillations are possible because the initial position of the kink corresponds not to the maximum itself, but to the point $x_0=-12$ located on the slope. The initial velocity of the kink $v=\Dot{x}_0(0)$ is zero. In the figure, $\varepsilon_1=0.1$ is assumed. Panel (a) shows the vibrations of the kink around its equilibrium position. As it slides toward the minimum, its speed increases, reaching a maximum value, and then decreases as it moves toward the adjacent maximum. The figure compares the trajectory obtained for the field solution (solid black line) with model 1 (dashed red line) and model 3 (dashed green line). As can be seen, the agreement between both effective models and the  solution of the field equation \eqref{phi4} in this case is striking and occurs for hundreds of time units (in the figure, at $t=400$, no deviations are visible).

Panel (b) compares the value of the variable $\gamma$ obtained from model 1 (red dotted line) with the values of $\gamma$ obtained from the solution of the field equation \eqref{phi4} (black dotted line). Visible here are the maxima of the variable $\gamma$, which correspond to the minima of the function $\mathcal{F}$.  It is worth noting that at this point the kink velocity is greatest. In turn, the turning points of the trajectory (shown in panel (a)) correspond to the minima of the function $\gamma$ in panel (b). At these points, the kink stops and the variable $\gamma$ takes the value $1$. The behavior of $\gamma$ suggests that, for the trajectory depicted in the figure, it admits a kinematic interpretation as the Lorentz factor.  It is worth emphasizing that in this approach, $\gamma$ is treated as a variable linking the kinematic component (the Lorentz factor) with the dynamic one.  For the variable $\gamma$, the agreement with the solution of the field equation \eqref{phi4} is exceptionally good.
 Furthermore, in panel (c), we compared the value of the variable  $b$ 
 obtained from model 3 (green dotted line) with the result of the field model  (black dotted line). Once again, the agreement is excellent.

\begin{figure}[h!]
    \centering
    \subfloat{{\includegraphics[height=4.2cm]{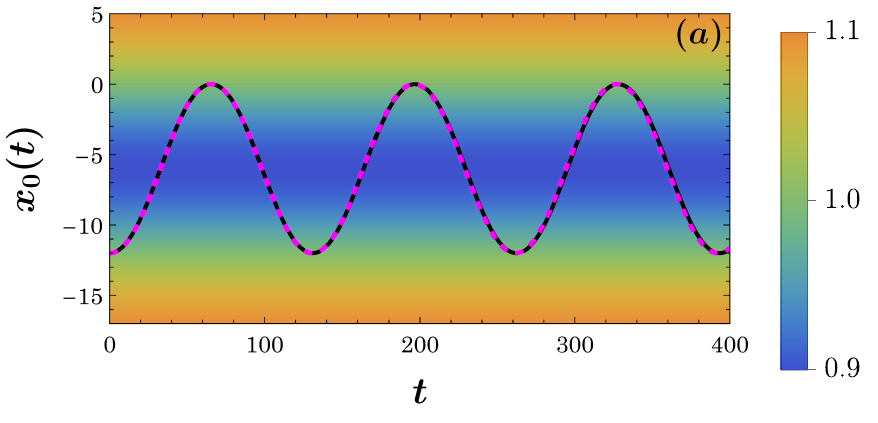}}}
    \quad
    \subfloat{{\includegraphics[height=4.5cm]{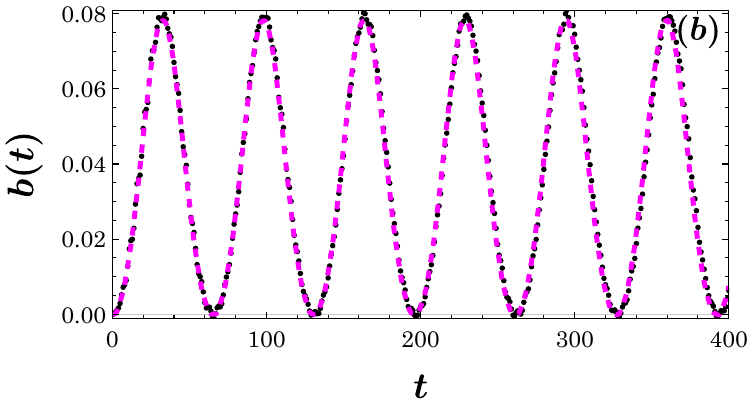}}}
    \caption{{Movement of a kink initially at rest ($v=0$) and located at $x_0=-12$. In this case, $\varepsilon_2=0$, so the system’s inhomogeneity arises solely from the function $\mathcal{F}$, with the parameter set to $\varepsilon_1=0.1$. (a) The kink trajectory computed from the field solution (black solid curve) is compared with that predicted by the second effective model (purple dashed curve). The color bar  on the right side of the
    panel indicates the values of the function $\mathcal{F}$. (b) The evolution of the variable $b$ obtained from the field equation \eqref{phi4} (black dotted curve) is compared with the corresponding result from the second effective model (purple dotted curve).}}
    \label{nfig_02}
\end{figure}

One might be interested in how model 2 would behave under the same conditions. Such a direct comparison is not possible due to the fact that within the framework of ansatz \eqref{kink-xi2++}, \eqref{xi2++} we are unable to reproduce the initial conditions \eqref{phi_wp1},\eqref{phi_wp2}, which was possible within the context of ansatz \eqref{kink-xi}, \eqref{xi}, defining model 1, and the ansatz \eqref{kink-xi2}, \eqref{xi2+}, defining model 3. In the context of model 2, we are forced to use the initial conditions \eqref{phi_wp-x}. For this reason, the results for model 2 are presented in a separate Figure \ref{nfig_02}. 

Panel (a) of this figure compares the positions of the kink $x_0$  obtained from the field equation \eqref{phi4} (solid black line in the figure) and  the same variable obtained from model 2 (purple dashed line). Clearly, the process shown in this figure is very similar to the process shown in Figure 1. This means that the kink, which initially rests at the point $x_0=-12$, slides down the “slope” (i.e., between the maximum and the minimum) while oscillating around the minimum of the function $\mathcal{F}$(x). On the right side of this panel is a color-bar indicating the values of the function $\mathcal{F}$. As in the previous figure, we have assumed $\varepsilon_2=0$ and $\varepsilon_1=0.1$.
On the other hand panel (b) compares the breathing mode vibrations obtained based on the field  model (dotted black line) and the effective model (dotted purple line). It is worth noting that in the case of such a simple evolution, the agreement between the two models for both variables $x_0$ and $b$ is excellent. 
 
\begin{figure}[h!]
    \centering
    \subfloat{{\includegraphics[height=4.2cm]{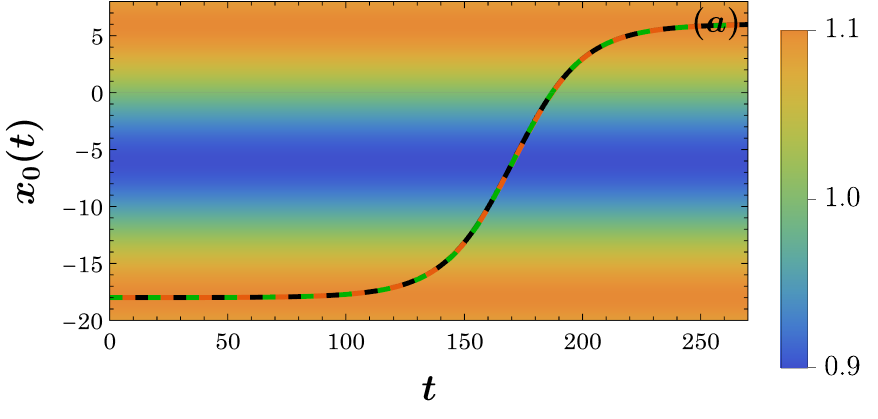}}}
    \quad
    \subfloat{{\includegraphics[height=4.5cm]{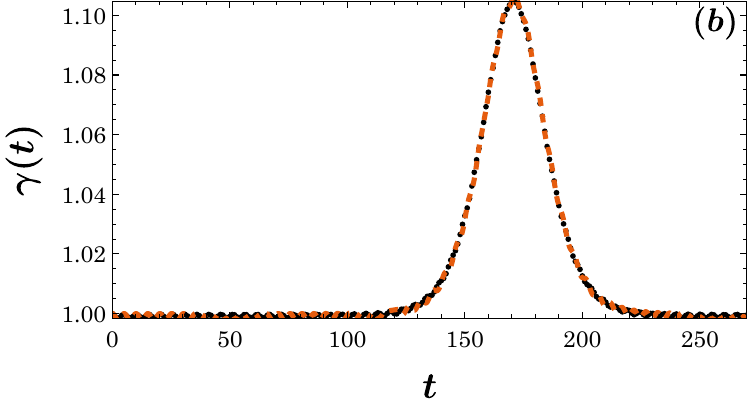}}}
    \quad
    \subfloat{{\includegraphics[height=4.5cm]{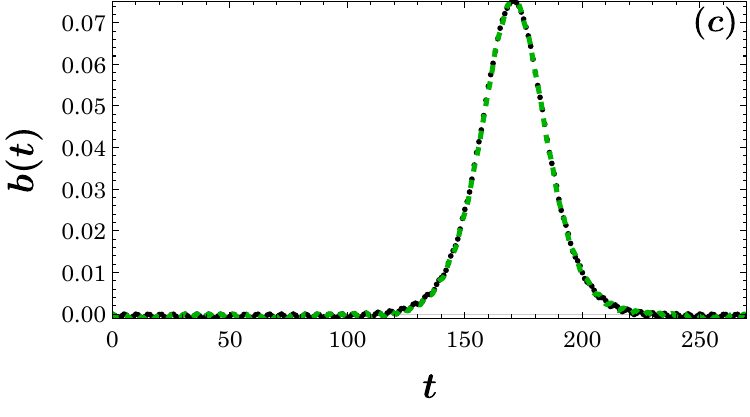}}}
    \caption{{Shifting the kink between two maxima of the function $\mathcal{F}$. As in the previous figure, $\varepsilon_1=0.1$ and $\varepsilon_2=0$. The initial position of the kink is $x_0=-18$. Initially, to enable the kink to slide, a minimum speed of $v=10^{-4}$ was assumed. (a) The continuous black line shows the trajectory obtained from the field equation \eqref{phi4}, the red dashed line shows the trajectory obtained from the first effective model, and the green dashed line was obtained from the third approximate model. The 
    color bar to the right of the panel refers to the values of the function $\mathcal{F}$. (b) Comparison of the variable $\gamma$ obtained from the field model (black dotted line) and the first approximate model (red dotted line). (c) The variable $b$ obtained from the third effective model (green dotted line) and the field equation  (black dotted line).}
 }
 \label{nfig_03}
\end{figure}

\break
Figure \ref{nfig_03} shows a similar situation, but this time the kink is initially set exactly at the maximum of the function ${\mathcal F}$, i.e., at position $x_0=-18$. To enable movement, the kink is given a slight initial velocity in the direction of the minimum. Panel (a) shows how the kink moves from one maximum to the next. The figure compares the trajectories obtained from three models: the field equation \eqref{phi4} (solid black line), the first effective model (dashed red line), and the third effective model (dashed green line). The results are in excellent agreement.
Panel (b) shows the dependence of the variable $\gamma$ present in the first approximate model on time. Like before, the predictions of the first effective model are represented by a red dotted line, while the result obtained from the field equation is represented by a black dotted line.
It can be seen that $\gamma$ changes from a value close to $\gamma=1$ to a value of $\gamma=1.1$. These results naturally correspond to the values of the Lorentz factor. 
A comparison of variable $b$ obtained from the third effective model (green dotted line) with the results of the field equation (black dotted line) is shown in panel (c). It can be seen that the compatibility for this model is as good as for the first model. 

\begin{figure}[h!]
    \centering
    \subfloat{{\includegraphics[height=4.2cm]{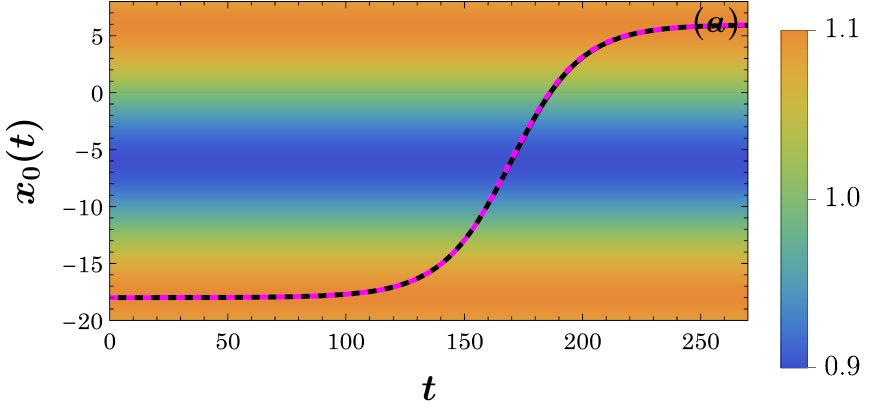}}}
    \quad
    \subfloat{{\includegraphics[height=4.5cm]{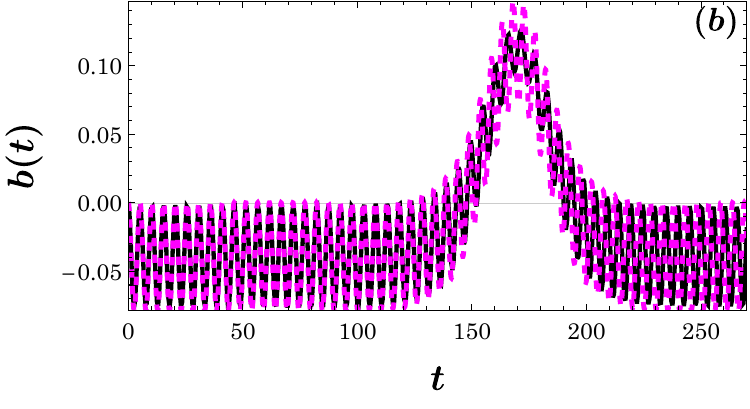}}}
    \caption{Displacement of the kink between two maxima of the function $\mathcal{F}$ for the initial conditions \eqref{phi_wp-x}. As in the previous figure, $\varepsilon_1=0.1$ and $\varepsilon_2=0$. The kink is initially located at $x_0=-18$. To allow it to move, a small initial velocity of $v=10^{-4}$ was introduced. (a) The solid black curve represents the trajectory obtained from the field equation \eqref{phi4}, while the purple dashed curve corresponds to the trajectory predicted by the second effective model. The color bar on the right indicates the values of the function $\mathcal{F}$. (b) Comparison of the variable $b$ as computed from the field equation (black dotted curve) and from the second approximate model (purple dotted curve).}
    \label{nfig_04}
\end{figure}

The situation shown in Figure \ref{nfig_03}  was reproduced in the case of model 2 for the initial conditions specified by equations \eqref{phi_wp-x}  in Figure \ref{nfig_04}. As in the previous figure, the kink moves from one maximum to the neighboring one. All parameters in the case described are identical to those in Figure \ref{nfig_03}. Panel (a) shows the kink trajectory. The solid black line shows the solution of the field equation \eqref{phi4}, while the dashed purple line was obtained based on model 2. The agreement between the two curves is very good here. As before, on the right side of the panel we have a color bar referring to the value of the $\mathcal{F}$ function. In the case of panel (b), we have a comparison of the variable $b$ describing the vibrations associated with the breathing mode in the solution of the field equation (black dotted line) and in the effective model 2 (purple dotted line).  Two phenomena can be observed in the figure. The first are vibrations of the field configuration. In fact, as we have noted, the occurrence of such vibrations is characteristic when we are dealing with an initial configuration that deviates from the configuration that minimizes the energy of the system.
 Width vibrations are a natural reaction of the system to the excess of energy trapped in it. 
Here ``excess energy'' refers to the energy localized around the kink. We determine it by taking the kink gradient to define a half-width, computing the energy density within this region for both models, and then comparing them—finding that model 2 yields about 1.6 \% higher energy.
 It is worth mentioning that the energy contained in the initial configuration \eqref{phi_wp1}, \eqref{phi_wp2} is so close to the minimum energy that such vibrations do not occur during evolution. Secondly, in this case, the variable $b$ obtained on the basis of the second effective model already shows some deviation from  the $b$ stemming from the field configuration.

\FloatBarrier
\subsubsection{Spatial modulation of $\mathcal{F}$  and $\lambda$}
The situation becomes more complicated when we assume a non-zero value for the parameter $\varepsilon_2$. If $\omega$ is equal to zero, then there is still no dependence of the coefficient $\lambda$ on time. In this case, the effective potential has a more complex form, as in Figure 3. In Figure \ref{nfig_05} the color bar on the right side of panel \ref{nfig_05} (a) refers to the effective potential.  Due to the more complex nature of the potential, the kink trajectory shown in panel (a) is modified in relation to Figure 1. In the simulation referring to the current figure, at the beginning of the evolution, the kink rested ($v=0$) at position $x_0=-16.5$. The parameter values in the figure are $\varepsilon_1=0.1$, $\varepsilon_2=0.05$ and $k=\pi/3$, respectively.
Note that the trajectory obtained from the first effective model (marked with a red dashed line) remains in perfect agreement with the trajectory obtained from the field {equation \eqref{phi4}(solid black line). The agreement extends to several hundred time units (in the figure, $t=300$). The situation is different for the third effective model. The trajectory obtained from this model (green dashed line), although similar in shape to the trajectory obtained from the field solution, begins to deviate significantly from it at $t=100$.
In the case under consideration, the second dynamic variable $\gamma$ has a very non-trivial behaviour. The time dependence of the variable $\gamma$  is shown in panel (b). 
In this figure, the red dotted line represents the outcome of the first effective model, while the black dotted line corresponds to the prediction obtained from the field equation \eqref{phi4}. 
Changes in the value of $\gamma$ are related largely to changes in the speed of the kink as it passes through local minima of the effective potential. Panel (c) compares the time dependence of variable $b$ of the third model (green dotted line) with the result of the field equation (black dotted line). The agreement between the two models applies to times limited to one hundred units. For longer times, the predictions of the approximate model gradually begin to deviate from the results obtained in the field model. 
\begin{figure}[h!]
    \centering
    \subfloat{{\includegraphics[height=4.2cm]{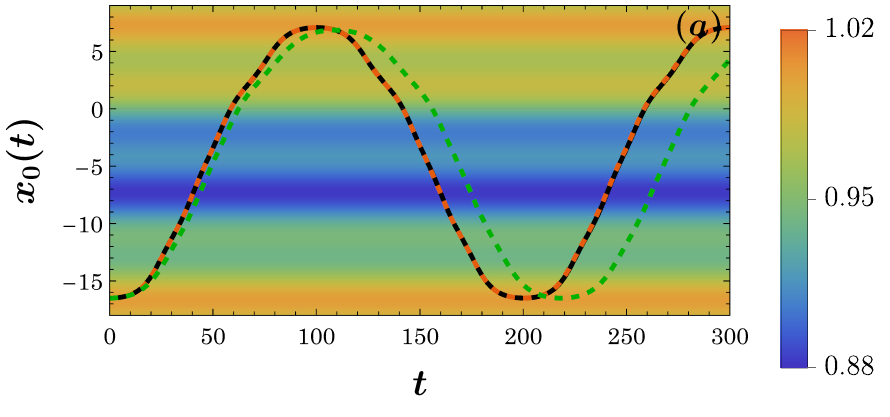}}}
    \quad
    \subfloat{{\includegraphics[height=4.5cm]{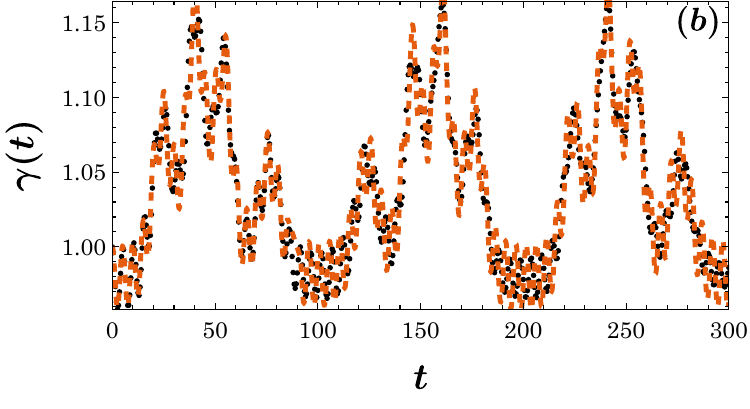}}}
    \quad
    \subfloat{{\includegraphics[height=4.5cm]{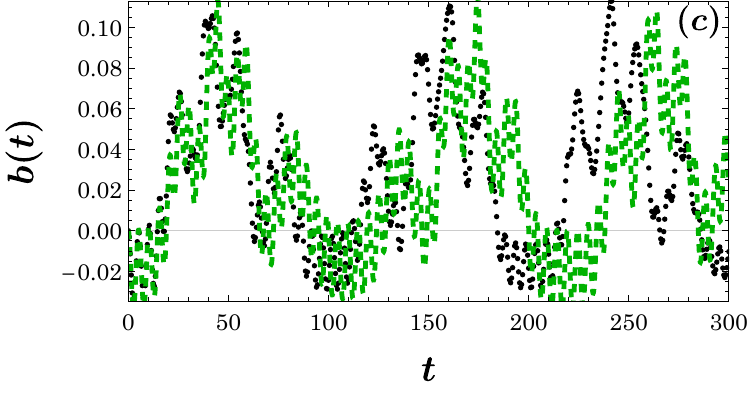}}}
    \caption{{Trajectory when $\varepsilon_1=0.1$ and $\varepsilon_2=0.05$. In this case, both $\mathcal{F}$ and $\lambda$ have a non-trivial shape. 
    This time, the color bar to the right of the first panel refers to the value of the effective potential.  The initial position of the kink is $x_0(0)=-16.5$, and the initial velocity is $v=0$. The parameter $k$ is equal to $\frac{\pi}{3}$. (a) Comparison of trajectories from three 
    computations: the field model \eqref{phi4} (shown by solid line), the first effective model (shown by dashed red line), and the third approximate model (shown by dashed green line). (b) The black dotted line shows the time dependence of the variable $\gamma$ in the field model, while the red dotted line shows the same variable obtained in the first effective model. (c) Again, the black dotted line shows the time course of variable $b$ obtained from the field model, while the green dotted line shows the result obtained from the third effective model.}
}
    \label{nfig_05}
\end{figure}

The calculations for the same parameter values as in Figure \ref{nfig_05} were performed for the initial conditions \eqref{phi_wp-x} and for model 2 (i.e., the more standard model). The results are shown in Figure \ref{nfig_06}.  The first surprise appears in panel (a), where the trajectory obtained based on the second effective model (purple dashed line) remains in fairly good agreement with the field model \eqref{phi4} (solid black line). This agreement is better than that of model 3, as can be seen by comparing Figure \ref{nfig_05} (a) with Figure \ref{nfig_06} (a). It is worth noting here that model 1 based on the variables $(x_0,\gamma)$ is by far the most accurate in this case. Panel (b) of Figure \ref{nfig_06}  shows the evolution of variable $b$, which is represented in the field model by a black dotted line and the same variable in the second effective model represented by a purple dotted line. As can be seen, the agreement between the two curves is reasonably good. This time, again due to the incompatibility of the initial conditions \eqref{phi_wp-x} to the energy-minimizing configuration, we observe fairly rapid oscillations.

\begin{figure}[h!]
    \centering
    \subfloat{{\includegraphics[height=4.2cm]{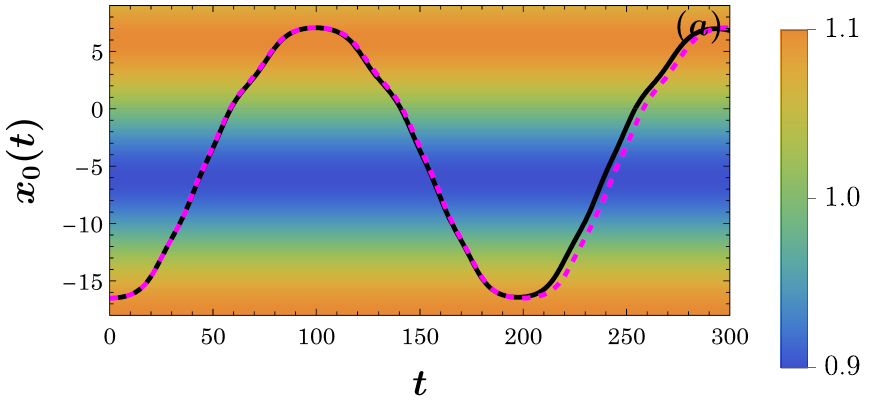}}}
    \quad
    \subfloat{{\includegraphics[height=4.5cm]{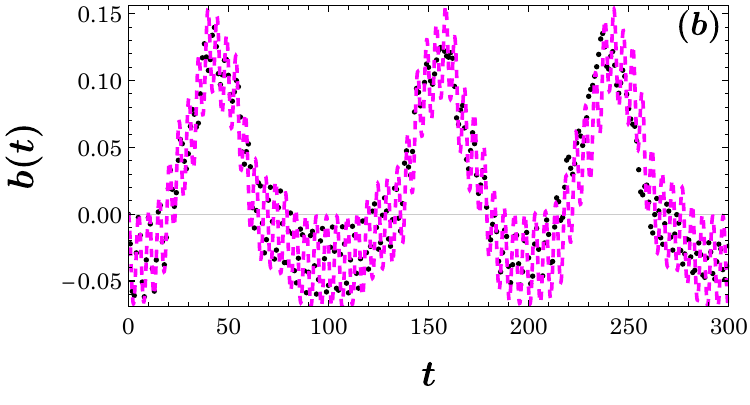}}}
    \caption{{Evolution of the kink for $\varepsilon_1=0.1$ and $\varepsilon_2=0.05$. The initial conditions \eqref{phi_wp-x} were adopted in the figure. In this setting, both $\mathcal{F}$ and $\lambda$ exhibit nontrivial spatial profiles. The color bar on the right of the first panel corresponds to the effective potential. The kink is initially placed at $x_0=-16.5$ with zero initial velocity ($v=0$), and the wavenumber is set to $k=\frac{\pi}{3}$. (a) The solid black line represents the trajectory derived from the field model \eqref{phi4}, while the dashed purple line shows the corresponding result from the second effective model. (b) Temporal evolution of the variable $b$ obtained from the field model (black dotted line) and from the second effective model (purple dotted line).}
}
        \label{nfig_06}
\end{figure}

\FloatBarrier
\subsection{Evolution in a non-autonomous system $\lambda(x,t)$ }
A significant complication introduced into the model is the assumption that $\varepsilon_2 \neq 0$ and $\omega \neq 0$. In this case, a non-trivial dependence on time appears in the function $\lambda$. First, we will present a scenario in which models are equally accurate.
Figure \ref{nfig_07} shows the process during which the kink, initially located
away from an equilibrium point, i.e., at $x_0=-6$ (with $v=0$),
begins to slide down the slope. It turns out that the temporal and spatial changes in the function $\lambda$, which result from non-zero values of the frequency ($\omega=0.08$) and $\varepsilon_2=0.1$, lead to the highly non-trivial trajectory course. The figure additionally assumes that $\varepsilon_1=0.4$ and $k=\frac{\pi}{6}$.
Panel (a) of this figure compares the trajectories obtained from three models: the original field model \eqref{phi4} (shown by black solid line) and two effective models. 
The red dashed line represents the outcome of the first approximate model, while the green dashed line corresponds to the prediction from the third effective model. Note that even for times $t=500$, the agreement between both approximate models and the field model is excellent. Similarly, panel (b) illustrates the close agreement between the time evolution of the dynamic variable $\gamma$ from the first effective model (red dotted line) and that obtained from the field model (black dotted line). The last panel (c) compares the time evolution of the dynamic variable $b$ of the third effective model (green dashed line) and the original field model (black dotted line). Also in this case, the similarity between the two models is remarkable, especially given the complex nature of the kink 
position variation.
\begin{figure}[h!]
    \centering
    \subfloat{{\includegraphics[height=4.5cm]{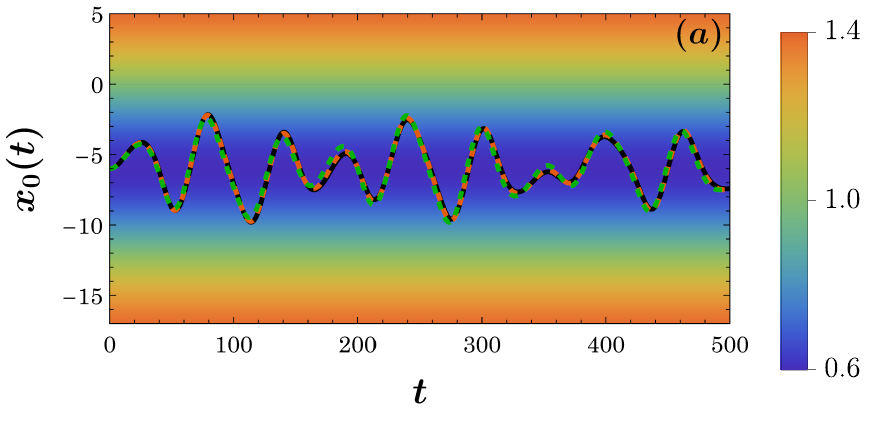}}}
    \quad
    \subfloat{{\includegraphics[height=4.5cm]{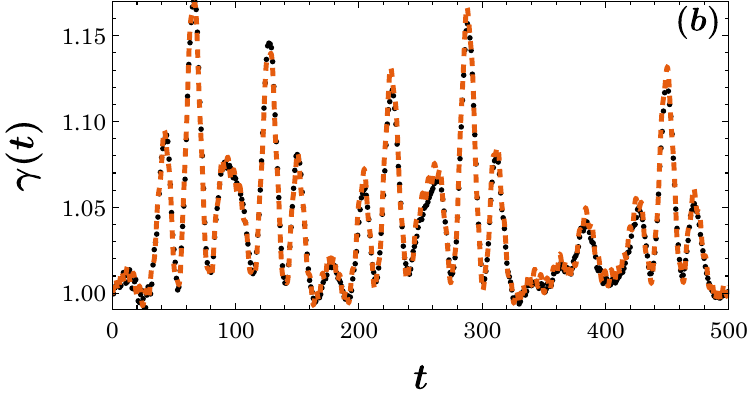}}}
    \quad
    \subfloat{{\includegraphics[height=4.5cm]{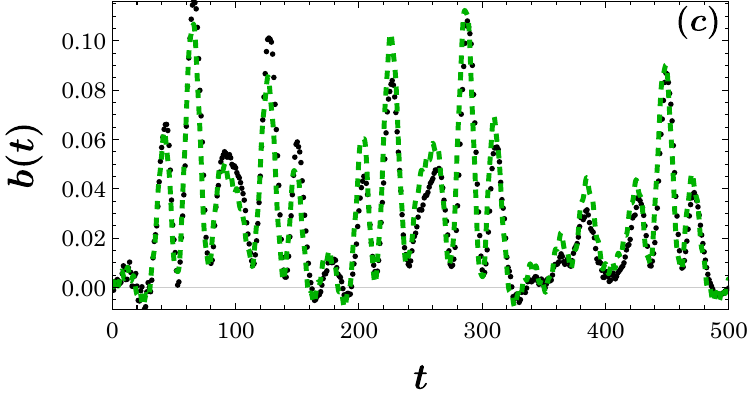}}}
    \caption{{Kink vibrations around the position $x_0=-6$ in the case of a non-autonomous system. The parameters in the figure are as follows: $\varepsilon_1 = 0.4$, $\varepsilon_2 = 0.1$, $k=\frac{\pi}{6}$, $\omega = 0.08$, $v=0$, and $x_0=-6$. (a) Comparison of the kink trajectory obtained from the field model \eqref{phi4} (shown by black solid line) with the trajectories obtained from two effective models: red dashed line for the first model, and green dashed line for the third model. As usual, the color bar on the right side of the panel refers to the values of the function $\mathcal{F}$. (b) Time course of the variable $\gamma$ obtained from the field model (black dotted line) and the first effective model (red dotted line). (c) Dependence of the variable $b$ on time in the field model (black dotted line) and the third effective model (green dotted line).}
}
    \label{nfig_07}
\end{figure}

The situation becomes more problematic when the results of the field model \eqref{phi4} are compared with those of the second effective model. Figure \ref{nfig_08} (a) shows the evolution under initial conditions \eqref{phi_wp-x}. 
As can be seen, the kink trajectory from the second effective model (purple dotted line) stays close to the black solid line corresponding to the field model, oscillating around it and crossing it multiple times.
This discrepancy probably has its origin in the presence of kink width vibrations, shown in panel (b), or rather in the transfer of energy between the translational and vibrational degrees of freedom. These excess vibrations in the field model have their origin in the choice of specific initial conditions.
 Comparing Figures \ref{nfig_07} (c) and \ref{nfig_08} (b), it can be seen that for the initial conditions \eqref{phi_wp-x}, these vibrations are much greater than in the former figure. Evidently, the deviations in the time dependence of variable $b$ in the second effective model (purple dotted line) and the field model (black dotted line) are very significant in the process under consideration. Nevertheless, it should be noted that despite these 
 substantial variations in $b$, the model does still proximally follow the gross
 features of the kink position evolution.

\begin{figure}[h!]
    \centering
    \subfloat{{\includegraphics[height=4.2cm]{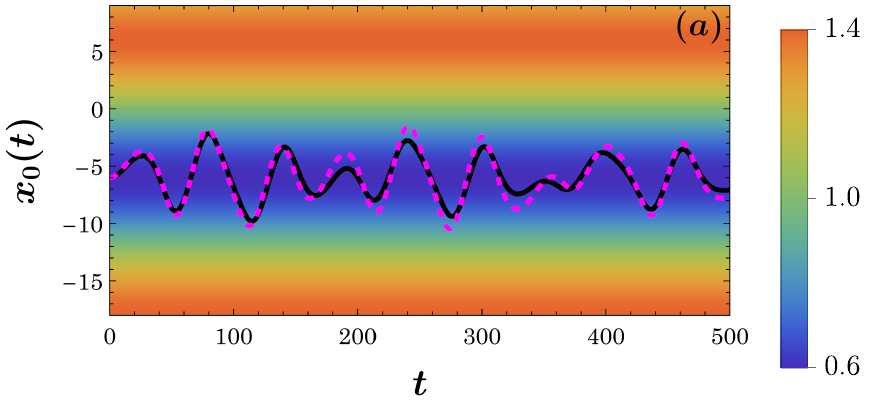}}}
    \quad
    \subfloat{{\includegraphics[height=4.5cm]{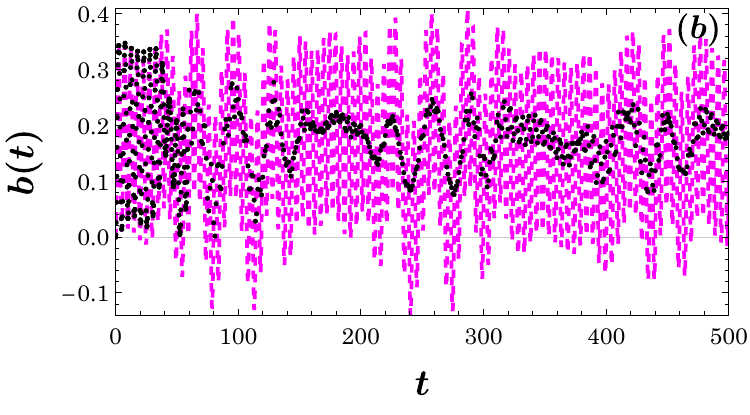}}}
    \caption{{Kink oscillations centered at position $x_0=-6$ in a non-autonomous system. The parameters used in the figure are: $\varepsilon_1 = 0.4$, $\varepsilon_2 = 0.1$, $k=\frac{\pi}{6}$, $\omega = 0.08$, $v=0$, and $x_0=-6$. (a) The black solid curve shows the kink trajectory computed from the field model \eqref{phi4}, while the purple dashed curve represents the trajectory predicted by the second effective model. 
    (b) Time evolution of the variable $b$ obtained from the field model (black dotted curve) and from the second effective model (purple dotted curve).}
    }
    \label{nfig_08}
\end{figure}

\break
The key parameter for validating both effective models is the frequency at which the magnitude of $\lambda$ varies - the very factor that introduces non-autonomy into the system under consideration. Figure \ref{nfig_09} shows how increasing the parameter to $\omega=0.04$ affects the agreement between the third effective model and the field model \eqref{phi4}. In the figure, the kink initially is placed (with $v=0$) at  $x_0=-6$. 
As in the previous case, $\varepsilon_1=0.4$ and $k=\frac{\pi}{6}$ are assumed, with $\varepsilon_2$ further specified as $0.15$ .
In panel (a), the black solid line represents the kink trajectory obtained from the field model \eqref{phi4}, i.e., the ``ground truth''. The red dashed line depicts the trajectory from the first approximate model, and the green dashed line corresponds to that from the third effective model. The color-bar on the right side of the figure refers to the values of the function $\mathcal{F}$. The figure shows that for all times (up to $t=600$), the first effective model remains consistent with the field model, while the consistency of the third model does not exceed $t=300$. Panel (b) of this figure compares the time course of variable $\gamma$ (red dotted line) with the result obtained based on the field model (black dotted line). It can be seen that up to time $t=500$, the two models agree very well, 
despite the multiple frequencies clearly involved within the kink
dynamics, while for longer times the agreement can still be considered acceptable.
Panel (c) confirms the effectiveness of the third model up to $t=300$. In the figure, the dependence of variable $b$ on time obtained on the basis of the field model is marked with a black dotted line, while on the basis of the effective model it is marked with a green dotted line. For times above $t=300$, the predictive power of the effective model becomes insufficient. 
{For Figures \ref{nfig_08} and \ref{nfig_09}, we also performed a Fourier analysis of both the signal defined by the position of the kink $x_0$ and the variable describing variations in the kink width (depending on the model, $b$ or $\gamma$). 
This analysis confirms a very good agreement between the spectra obtained from models 1 and 3 and the corresponding spectrum of the field model. In contrast, for model 2 a small but noticeable frequency shift appears in the variable $x_0$. This observation is 
reflected in the dynamics of Figure \ref{nfig_08}.}

\begin{figure}[h!]
    \centering
    \subfloat{{\includegraphics[height=4.5cm]{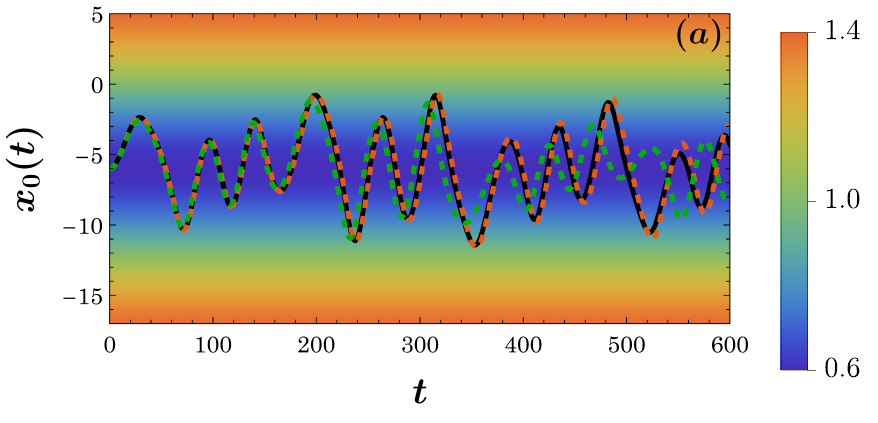}}}
    \quad
    \subfloat{{\includegraphics[height=4.5cm]{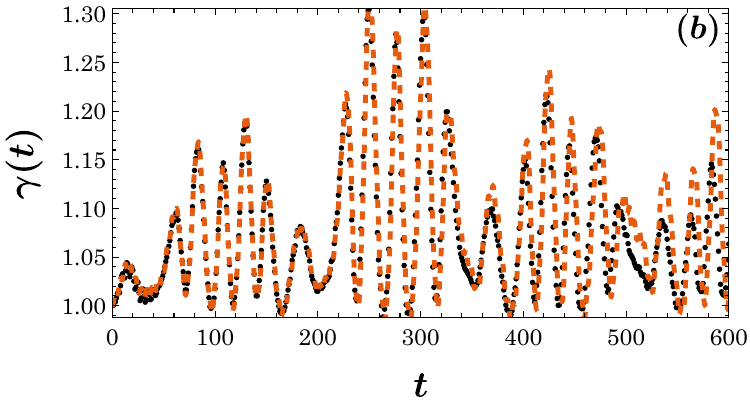}}}
    \quad
    \subfloat{{\includegraphics[height=4.5cm]{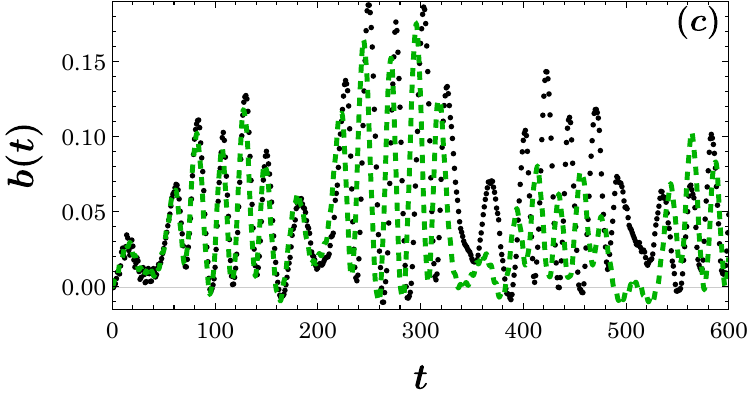}}}
    \caption{Vibrations as in the prior figure, in the case of a non-autonomous system for a different set of parameters. In the figure, the following assumptions were made: $\varepsilon_1 = 0.4$, $\varepsilon_2 = 0.15$, $k=\frac{\pi}{6}$, $\omega = 0.04$, $v=0$, and $x_0=-6$. (a) Comparison of the kink trajectory obtained from the field model \eqref{phi4} (black solid line) and the trajectories obtained from both effective models: red dashed line (first model), green dashed line (third model). The color-bar refers to the values of the function $\mathcal{F}$. (b) Dependence of the variable $\gamma$ on time obtained from the field model (black dotted line) and the first effective model (red dotted line). (c) Time course of the variable $b$ in the field model (black dotted line) and the third approximate model (green dotted line).}
    \label{nfig_09}
\end{figure}

For comparison, the predictions of the standard effective model (model 2) for the same parameter values and initial conditions \eqref{phi_wp-x} are shown in Figure \ref{nfig_10}. Note that in the case of variable $x_0$, whose time dependence is shown in panel (a), there is a significant difference between the result obtained on the basis of the field model \eqref{phi4} (black solid line) and that obtained on the basis of the second effective model (purple dashed line).   The trajectory of the effective model deviates from the trajectory obtained within the framework of the field model already for times slightly above 100 units. Panel (b) of this figure also shows significant differences for variable $b$. The result obtained from the effective model, represented by the purple dotted line, deviates significantly from the field model result represented by the black dotted line. Indeed, here, the dynamics of the kink motion are
accurately captured only for early times. In the vicinity of $t=100$, the
dynamics of $b$ is no longer accurately captured, eventually building up
(at longer times) an increasing deviation in the dynamics of $x_0(t)$.
\begin{figure}[h!]
    \centering
    \subfloat{{\includegraphics[height=4.2cm]{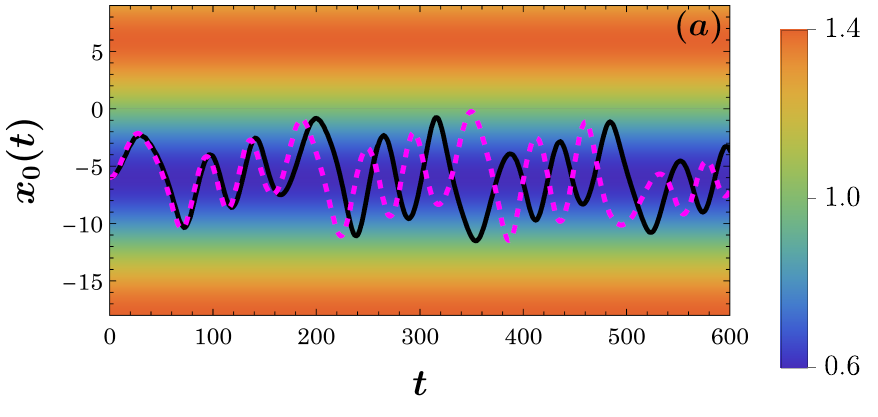}}}
    \quad
    \subfloat{{\includegraphics[height=4.5cm]{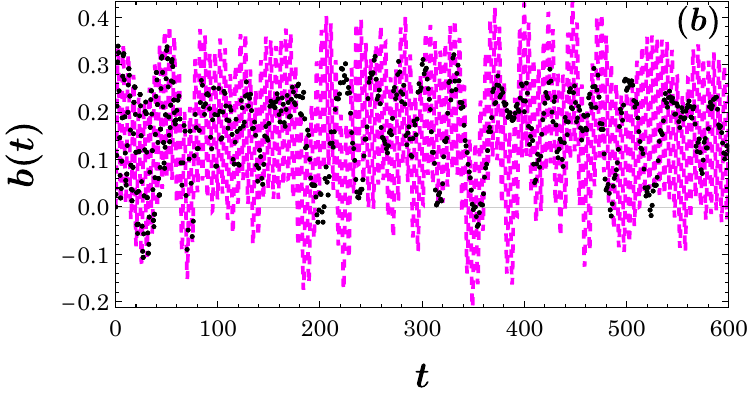}}}
    \caption{Vibrations in a non-autonomous system.  The parameters are the same as in the previous figure i.e.: $\varepsilon_1 = 0.4$, $\varepsilon_2 = 0.15$, $k = \frac{\pi}{6}$, $\omega = 0.04$, $v = 0$, and $x_0 = -6$. (a) Comparison between the kink trajectory obtained from the field model \eqref{phi4} (black solid line) and that derived from the second effective model (purple dashed line).
    (b) Time evolution of the variable $b$ obtained from the field model (black dotted line) and the second effective model (purple dotted line).}
    \label{nfig_10}
\end{figure}

A similar situation occurs in Figure \ref{nfig_11}. Compared to the previous figure, the modulation of $\mathcal{F}$, $\varepsilon_1$ has been reduced to $0.1$. Both the $\lambda$ parameters $\varepsilon_2=0.15$ and $k=\frac{\pi}{6}$ are identical to those in the previous figure. In addition, the kink begins its evolution at point $x_0=-6$, where it initially rests, i.e., $v=0$. The frequency in the function $\lambda$ has been increased to $\omega=0.05$.
Panel (a) of this figure, as in the previous cases, compares the kink trajectory from the field model \eqref{phi4} (solid black line) with those obtained from the first (dashed red line) and third  (dashed green line) effective models. The color-bar on the right-hand side of the figure lists the corresponding values of the function~$\mathcal{F}$.
Note that, while the first model remains consistent over the time interval shown in the figure, the third model exhibits a noticeable deviation already at $t=70$. Panel (b) shows excellent agreement between the variable $\gamma$ determined based on the first approximate model (red dotted line) and that determined based on the field model (black dotted line). The last panel (c) shows that the variable $b$ obtained from the third approximate model (green dotted line) agrees with the field model only up to $t=50$. For longer times, the green dotted line deviates quite significantly from the black dotted line determined based on the field model.
\begin{figure}[h!]
    \centering
    \subfloat{{\includegraphics[height=4.5cm]{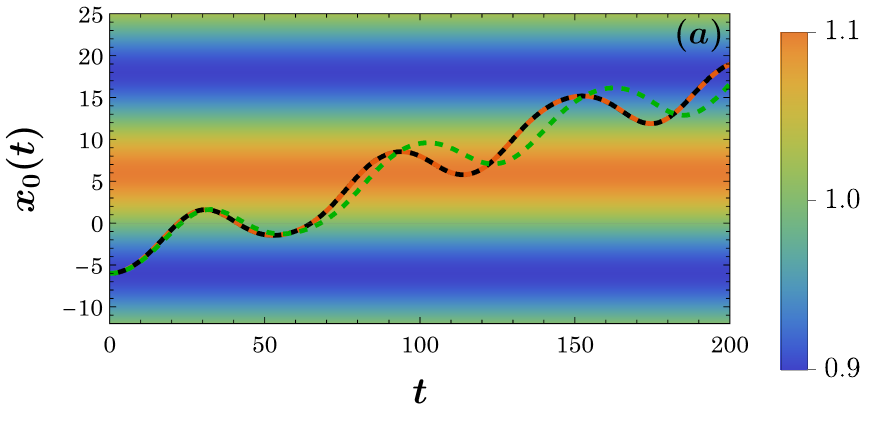}}}
    \quad
    \subfloat{{\includegraphics[height=4.5cm]{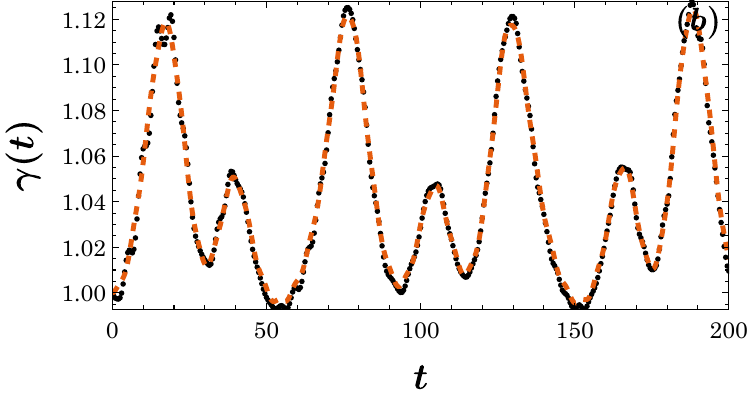}}}
    \quad
    \subfloat{{\includegraphics[height=4.5cm]{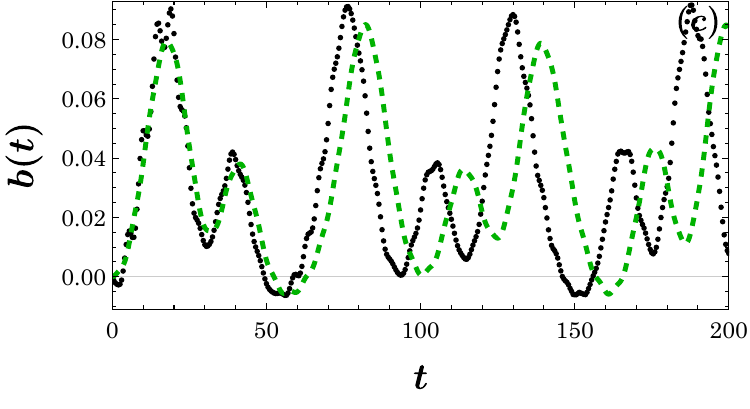}}}
    \caption{{Transport of the kink through the maximum of the function $\mathcal{F}$, resulting from the time dependence of the function $\lambda$. The following values were assumed in the figure: $\varepsilon_1 = 0.1$, $\varepsilon_2 = 0.15$, $k=\frac{\pi}{6}$, $\omega = 0.05$, $v=0$, and $x_0=-6$. (a) The kink trajectory obtained from the field model \eqref{phi4} (black solid line) compared with the trajectories obtained from the effective models: red dashed line (first model), green dashed line (third model).
    (b) The dependence of the variable $\gamma$ on time, as obtained within the framework of the field model,  is represented  by the black dotted line and is compared to the red dotted line representing the result obtained in the effective model. (c) The dependence of the variable $b$ on time in the field model (black dotted line) and the third approximate model (green dotted line).}}
    \label{nfig_11}
\end{figure}

The transmission of the kink through the barrier is also illustrated in Figure \ref{nfig_12}. This figure compares the second effective model with the field model \eqref{phi4} for the same set of parameters as in Figure \ref{nfig_11}. In this case, the field model employs the initial conditions \eqref{phi_wp-x}, which are consistent with the ansatz \eqref{kink-xi2++}, \eqref{xi2++}. Panel (a) presents the trajectory obtained from the field model (black solid line) alongside that from model 2 (purple dashed line). The two trajectories exhibit excellent agreement. The excess energy present in the initial configuration induces oscillations of the variable $b$, as shown in panel (b). In this panel, the black dotted line corresponds to the field model, while the purple dotted line represents model 2. Since the frequency of these oscillations is close to that of the breathing mode, it appears that the variable $b$ absorbed the excess energy, allowing the variable $x_0$ to remain unaffected and thus to reproduce the field trajectory accurately (as seen in panel (a)). In this respect, Figure \ref{nfig_12} (a) differs from Figure \ref{nfig_10} (a) or \ref{nfig_08} (a), where the variable $x_0$ 
apparently absorbs only part of the excess energy.

\begin{figure}[h!]
    \centering
    \subfloat{{\includegraphics[height=4.2cm]{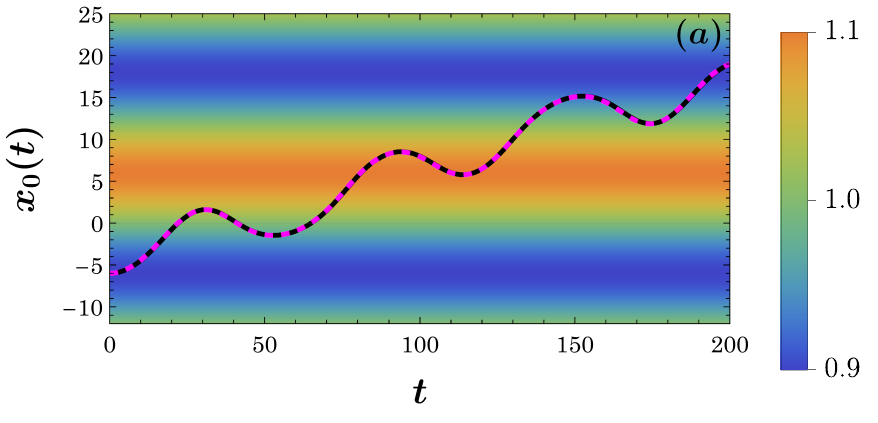}}}
    \quad
    \subfloat{{\includegraphics[height=4.5cm]{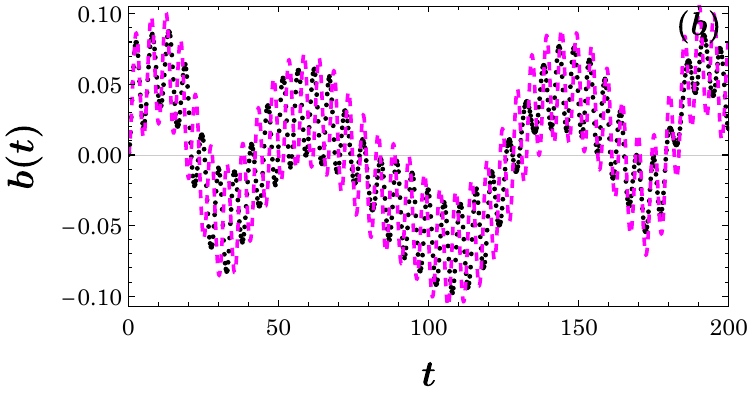}}}
    \caption{{The figure was generated for the following parameter values: $\varepsilon_1 = 0.1$, $\varepsilon_2 = 0.15$, $k = \frac{\pi}{6}$, $\omega = 0.05$, $v = 0$, and $x_0 = -6$. (a) Trajectory of the kink obtained from the field model \eqref{phi4} (black solid line) compared with that predicted by the second effective model (purple dashed line).  
    (b) Time evolution of the variable $b$ in the field model (black dotted line) and in the second effective model (purple dotted line).}}
    \label{nfig_12}
\end{figure}

We also test the models in the regime of substantially increased frequency. The predictive power of the effective models is preserved, as demonstrated in Fig.~\ref{nfig_13}. In both panels, the kink is initially placed at the point $x_0=-5$ with vanishing velocity, i.e., $v=0$. The remaining parameters are fixed to $\varepsilon_1 = 0.1$ and $k=\frac{\pi}{6}$. In both cases, the kink evolves within a single minimum of the function $\mathcal{F}$, as indicated by the black curve obtained from the field model \eqref{phi4}. It is worth noting that, in order to maintain agreement between the effective models and the field description, we had to keep $\varepsilon_2$ at a relatively low level. Panel (a) shows the trajectories for $\omega=1$, with $\varepsilon_2=0.15$. At this frequency, the first effective model (red dashed line) remains consistent with the field model, exhibiting good agreement up to times of order $t=200$. A similar behavior is observed for the third effective model (green dashed line), although its accuracy is slightly lower. Panel (b) corresponds to a much higher frequency, $\omega=10$, and $\varepsilon_2=0.05$. In this case, the first effective model (red dashed line) again remains consistent with the field model at least up to times of order $t=200$. The third effective model (green dashed line) achieves a comparable level of accuracy. It is worth noting that in the situation illustrated in panel (b), the kink trajectory remains smooth. This behavior arises because the small forces acting frequently and in alternating manner effectively average out, leaving the kink no time to respond to individual perturbations. In contrast, in panel (a) the forces occur at a much lower frequency, allowing the kink to react more noticeably, which manifests as small oscillatory structure in its trajectory.

\begin{figure}[h!]
    \centering
    \subfloat{{\includegraphics[height=3.9cm]{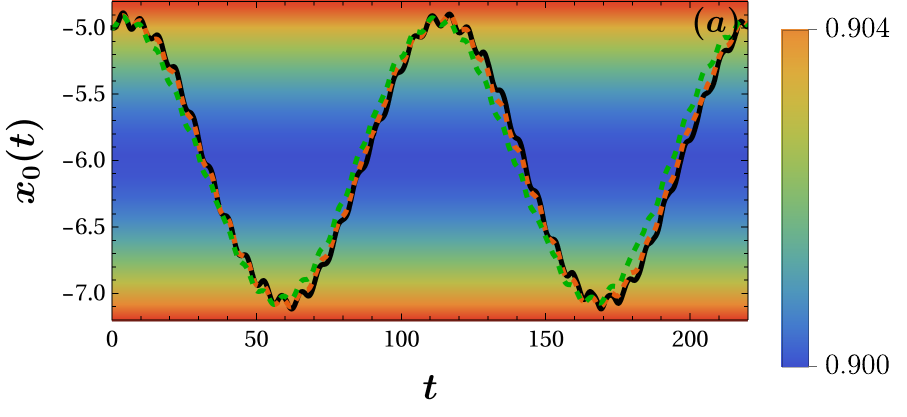}}}
    \quad
    \subfloat{{\includegraphics[height=3.9cm]{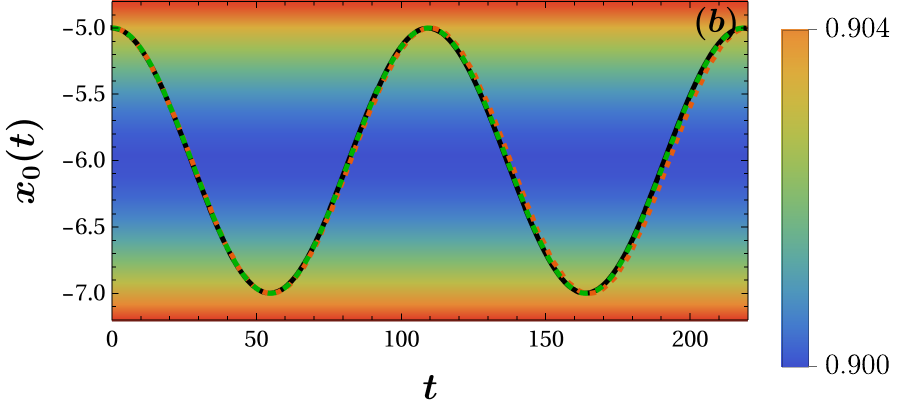}}}
    \caption{ Kink trajectory in the high-frequency forcing regime, compared with the predictions of the first and third models.  The following parameters were used in both figures: $\varepsilon_1 = 0.1$, $k=\frac{\pi}{6}$, $v=0$, and $x_0=-5$. 
    As in the previous figures, the black line corresponds to the results of the field model \eqref{phi4}, while the red dashed line represents the results of the first model, and the green dashed line those of the third model.
    The frequency values in both figures are as follows: (a) $\varepsilon_2 = 0.15$ and $\omega = 1.0$, (b) $\varepsilon_2 = 0.05$ and $\omega = 10.0$.}
    \label{nfig_13}
\end{figure}

Figure~\ref{nfig_14} illustrates the dynamics of the kink in the second effective model for the same set of parameters as in Fig.~\ref{nfig_13}. Panels (a) and (b) differ in the frequencies entering the function $\lambda$ as well as in the value of $\varepsilon_2$. In both panels, the black solid curve corresponds to the trajectory obtained from the field model \eqref{phi4}, whereas the purple dashed curve represents the prediction of the second effective model. The color bars on the right-hand side of each panel indicate the values of the function $\mathcal{F}(x)$. The level of agreement observed in panels (a) and (b) is comparable to that achieved by the first effective model in Fig.~\ref{nfig_13}.

\begin{figure}[h!]
    \centering
    \subfloat{{\includegraphics[height=3.9cm]{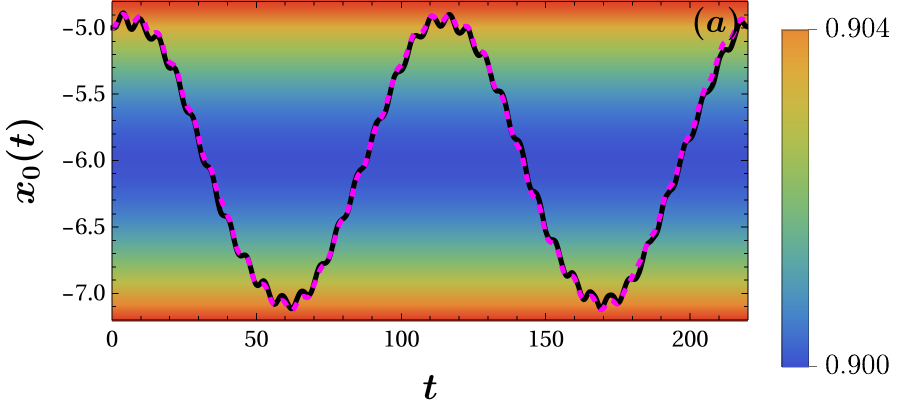}}}
    \quad
    \subfloat{{\includegraphics[height=3.9cm]{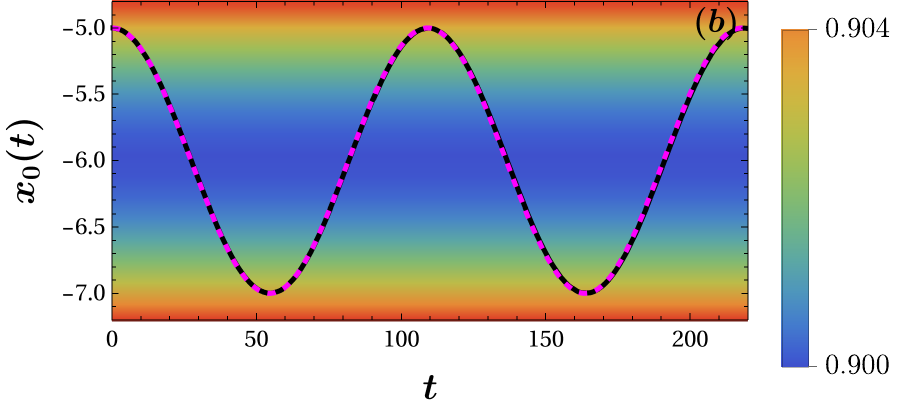}}}
    \caption{Kink trajectory compared with the results of the second model in the high-frequency perturbation regime. (b). Both panels were generated using the following parameter values: $\varepsilon_1 = 0.1$, $k = \frac{\pi}{6}$, $v = 0$, and $x_0 = -5$. A reference bar on the right-hand side of the figure indicates the corresponding values of the function $\mathcal{F}$. As in the previous figures, the black solid line represents the results obtained from the field model \eqref{phi4}, while the purple dashed line corresponds to the predictions of the second effective model. The frequency values are: (a) $\varepsilon_2 = 0.15$ and $\omega = 1.0$, (b) $\varepsilon_2 = 0.05$ and $\omega = 10.0$.}
    \label{nfig_14}
\end{figure}

The results presented above using typical trajectories obtained during the study clearly show that the model based on variables $(x_0,\gamma)$, i.e., model 1, most adequately reflects the field trajectory. This conclusion also holds when the disturbance is introduced via the time dependence of the function $\lambda$, particularly in the high-frequency regime. The typical examples presented in this section show very good agreement between model 1 and the field model {\color{red} \eqref{phi4}} (both for the
collective coordinate $x_0(t)$ and for the variable $\gamma(t)$). On the other hand, models 2 and 3 are definitely less well suited to the trajectory obtained based on the field model. The reason for the deviations of models 2 and 3 from the trajectory described by model 1 can be attributed to the deviations in the shape of the vibration modes present in models 2 and 3 from the shape of the Derrick mode, which is an integral part of model 1; we explain this in
more detail in 
\hyperref[AppendixC]{Appendix C}. 
We understand the Derrick mode as the term that appears at first order in the expansion of the ansatz (on the basis of which model 1 was constructed) with respect to the variable $\gamma$ around unity. We compare this mode to the vibrational modes appearing in the ans{\"a}tze used to construct both models 2 and 3. We adopt this comparison strategy because model 1 exhibits, in all cases considered, the best overall agreement with the field model. It turns out, however, that in some of the cases analyzed here model 2, and in others model 3, provides a better fit to the trajectories obtained from the field model. This sentence naturally refers only to the relative accuracy of models 2 and 3 compared to the field model. We find that whenever the vibrational mode of a given model more accurately approximates the Derrick mode, that model (either 2 or 3) reproduces the results of the field model more faithfully (and, naturally, also those of model 1).

\FloatBarrier
\section{The system with dissipation}
\label{sec5}
Since models 2 and 3 describe the field model \eqref{phi4} less accurately than model 1, we will only describe the dissipative system using model 1 in what follows. 
\subsection{Effective model}
One possible framework for describing dynamical systems in the presence of dissipation is the non-conservative variational approach \cite{Galley2013,Kevrekidis2014}. In its traditional, conservative form, the variational principle is posed as a boundary-value problem in time: the dynamical variables (or fields) are fixed both at the initial and the final instants of the evolution. This construction is well suited for reversible dynamics, but it inherently excludes genuinely irreversible processes, since the forward and backward time evolutions are treated on equal footing. To overcome this limitation, Galley’s formulation \cite{Galley2013} modified the very structure of the variational principle. Instead of prescribing the values of the dynamical variables at both temporal boundaries, only the initial configuration is fixed; the variables at the final time remain unconstrained. This relaxation of the final-time condition is not done directly but is achieved through a doubling of the dynamical variables: each physical degree of freedom is replaced by a pair of copies, evolving along two independent “histories.” By carefully constructing an extended Lagrangian that depends on both copies, one can set up a variational problem in which the variations at the initial time are fixed, while those at the final time are left completely free.
In the last step of the procedure-referred to as taking the physical limit-the two copies are identified with each other, thereby recovering the actual physical trajectory of the system. The advantage of this doubled-variable formalism is that it naturally accommodates dissipative effects, since the asymmetry between the initial and final boundaries allows for the description of processes in which energy is not conserved.
Formally, the Lagrangian density in this non-conservative variational framework is written as:
\begin{equation}
    \label{nonconservative}
    {\cal L}_N = {\cal L}(\phi_1) - {\cal L}(\phi_2) + {\cal R} .
\end{equation}
Within this framework, the resulting equations of motion retain a structure closely resembling that of the conventional, conservative formulation. The essential distinction lies in the appearance of an additional term on the right-hand side, which explicitly encodes the influence of dissipative mechanisms present in the system. This extra contribution modifies the familiar form of the dynamics, ensuring that energy loss or other non-conservative effects are properly incorporated into the evolution
\begin{equation}
    \label{eq-nonconservative}
    \partial_{\mu} \left( \frac{\partial {\cal L}}{\partial (\partial_{\mu} \phi)}\right) - \frac{\partial {\cal L}}{\partial \phi} = \left[\frac{\partial {\cal R}}{\partial \phi_{-}} - \partial_{\mu} \left( \frac{\partial {\cal R}}{\partial (\partial_{\mu} \phi_{-})}\right)\right]_{PL},
\end{equation}
where the index $\mu$ enumerates the space-time variables
$x^{\mu}=(x^0,x^1)=(t,x)$ and we assumed the standard summation
convention. Moreover we used variables, $\phi_{-}$ as well as
$\phi_{+}$ which are related to the original variables $\phi_1$
and $\phi_2$ as follows: $\phi_1=\phi_{+}+ \frac{1}{2}\phi_{-}$
and $\phi_2=\phi_{+}- \frac{1}{2}\phi_{-}$ (or conversely
$\phi_+=(\phi_1+\phi_2)/2$ and $\phi_-=\phi_1-\phi_2$).
Additionally, PL denotes the physical limit, in
which $\phi_{+}$ becomes a physical variable $\phi_{+}=\phi$ and
$\phi_{-}=0$. For the Lagrangian density \eqref{L} considered in
this article, the equation \eqref{eq-nonconservative} can be
converted to the form
\begin{equation}
\label{eq-nonconservative2}
    \partial_t^2 \phi - \partial_x (\mathcal{F}(x)\partial_x \phi) + \lambda(t,x) \, \phi (\phi^2 -1) = \left[\frac{\partial {\cal R}} {\partial \phi_{-}} - \partial_{\mu} \left( \frac{\partial {\cal R}}{\partial (\partial_{\mu} \phi_{-})}\right)\right]_{PL} .
\end{equation}
In Galley’s formalism, the dissipative dynamics are encoded precisely in the evolution of $\phi_{-}$, which vanishes in the absence of non-conservative effects. Working in the $(\phi_{+},\phi_{-})$ basis therefore simplifies both the variational procedure and the identification of dissipative contributions. 
We choose the non-conservative part of the Lagrangian $\mathcal{R}$ so that equation \eqref{eq-nonconservative2} reproduces the result of Eq. \eqref{phi4+}
\begin{equation}
   {\cal R} = -\Gamma \phi_{-} - \eta \phi_{-} \partial_t \phi_{+} .
\end{equation}
Starting from the non-conservative part of the Lagrangian, 
we proceed by substituting the ansatz given in Eqs. \eqref{kink-xi} and \eqref{xi} into the corresponding expression for $\mathcal{R}$.
This substitution yields the explicit dependence of the non-conservative contribution on the collective coordinate(s) introduced through the ansatz. In order to reduce the resulting expression to an effective, two-dimensional description, we perform an integration over the spatial coordinate,
 $$R_{eff} = \int_{-\infty}^{+\infty} dx {\cal R} . $$ 
The quantity $\mathcal{R}_{eff}$  thus represents the net effect of the non-conservative term on the collective-coordinate dynamics after averaging over the spatial profile of the field configuration.  With this effective contribution at hand, the equations of motion for the reduced system can be cast in the  form
\begin{equation}
    \label{eff-eq1}
    \frac{d}{d t }\left(\frac{\partial L_{eff}}{\partial \Dot{x}_0} \right) - \frac{\partial L_{eff}}{\partial {x}_0} = \left[ \frac{\partial R_{eff}}{\partial {x}_{-}} -
    \frac{d}{d t }\left(\frac{\partial R_{eff}}{\partial \Dot{x}_{-}} \right) \right]_{PL} ,
\end{equation}
\begin{equation}
    \label{eff-eq2}
    \frac{d}{d t }\left(\frac{\partial L_{eff}}{\partial \Dot{\gamma}} \right) - \frac{\partial L_{eff}}{\partial {\gamma}} = \left[ \frac{\partial R_{eff}}{\partial {\gamma}_{-}} -
    \frac{d}{d t }\left(\frac{\partial R_{eff}}{\partial \Dot{\gamma}_{-}} \right) \right]_{PL} ,
\end{equation}
The effective Lagrangian $L_{eff}$, whose explicit form was derived in the preceding section Eq. \eqref{Leff}, now serves as the starting point for the reduced description. 
In the present context we focus on the physical limit applied to the reduced collective coordinates $x_0$ and $\gamma$. In this limit, the “difference” variables vanish $x_{-}=0$, $\gamma_{-}=0$, while the “average” variables are identified with the physical degrees of freedom, $x_{+}=x_0$ and $\gamma_{+}=\gamma$. 
It is worth noting that the transformation between the doubled set of reduced variables and their $(+,-)$ counterparts is completely analogous to the corresponding relations for the field variables discussed earlier. Explicitly, we have $x_1=x_+ + \frac{1}{2} x_-$, $x_2=x_+ - \frac{1}{2} x_-$ and $\gamma_1=\gamma_+ + \frac{1}{2} \gamma_-$, $\gamma_2=\gamma_+ - \frac{1}{2} \gamma_-$. 
This parallel structure ensures that the passage from the field description to the reduced, collective-coordinate framework preserves the same symmetry between “average” and “difference” variables, thereby maintaining consistency with the non-conservative variational formalism introduced above.
We note in passing that in the context of such Klein-Gordon field models, this collective coordinate formulation of non-conservative
settings was introduced 
in the work of~\cite{Kevrekidis2014}.
Effective equations of motion \eqref{eff-eq1}, \eqref{eff-eq2} for a model with two degrees of freedom therefore assume the final form
\begin{equation}
\begin{gathered}
\label{2dof_ansatz_dis}
    M\Ddot{x}_0+\kappa \Ddot{\gamma}+ (\partial_{\gamma} M) \Dot{x}_0 \Dot{\gamma}+\frac{1}{2}(\partial_{x_0}M)\Dot{x}_0^2+\left(\partial_{\gamma} \kappa-\frac{1}{2}\partial_{x_0}m \right)\Dot{\gamma}^2+ \\
    (\partial_t M) \Dot{x}_0 +\left(\partial_{\gamma} \alpha + \partial_t \kappa -\partial_{x_0} \beta \right) \Dot{\gamma} + \partial_t \alpha +  \partial_{x_0}V=\\
2 \Gamma - \eta \frac{1}{\gamma} \sqrt{\frac{2 {\mathcal{F}}}{\lambda}} \left\{ c \left(\frac{\partial_{x_0} \lambda}{ 2 \lambda} - \frac{\partial_{x_0} \mathcal{F}}{2 \mathcal{F}}\right)  \left[ \frac{\partial_t \lambda}{ 2 \lambda} + \frac{\Dot{\gamma}}{\gamma} + \left( \frac{\partial_{x_0} \lambda}{2 \lambda} - \frac{\partial_{x_0} \mathcal{F}}{2 \mathcal{F} }\right) \Dot{x}_0 \right] 
+ \frac{2 \lambda}{3 \mathcal{F}} \gamma \Dot{x}_0
\right\}
    \\
    m\Ddot{\gamma} + \kappa \Ddot{x}_0+ (\partial_{x_0} m) \Dot{x}_0 \Dot{\gamma}+\frac{1}{2}(\partial_{\gamma}m)\Dot{\gamma}^2
    + \left( \partial_{x_0} \kappa
    -\frac{1}{2}\partial_{\gamma}M \right)\Dot{x}_0^2 +\\
(\partial_t m) \Dot{\gamma} + \left(\partial_{x_0} \beta + \partial_t \kappa - \partial_{\gamma} \alpha \right) \Dot{x}_0  +\partial_t \beta  +   \partial_{\gamma}V=\\
 - \eta \frac{1}{\gamma^2} \sqrt{\frac{2 {\mathcal{F}}}{\lambda}} ~c \left[  \frac{\partial_t \lambda}{ 2 \lambda} + \frac{\Dot{\gamma}}{\gamma} + \left( \frac{\partial_{x_0} \lambda}{2 \lambda} - \frac{\partial_{x_0} \mathcal{F}}{2 \mathcal{F} }\right) \Dot{x}_0 \right] 
.
\end{gathered}
\end{equation}
The constant $c$ appearing in the above equations takes the value
$c=\frac{\pi^2}{9}-\frac{2}{3}$. All the necessary coefficients and detailed expressions required for implementation of the presented model are provided in \hyperref[AppendixA]{Appendix A}, specifically in the formulas labeled as equation \eqref{integrals+c}. 
A natural question concerns the usefulness of such rather elaborate
effective models. It is important to note in that regard that these
models are still explicitly analytical in their form, and as such
they enable a multitude of relevant analyses, including the identification
of stationary states, and the consideration of stability around such
stationary states in a practically explicit form. They are also 
immensely faster to simulate and obtain information from in comparison
to PDE dynamics (indeed even more so in higher-dimensional settings).
Hence, we feel that despite their additional complexity such models
are still highly valuable from both an analytical and a numerical
perspective.

\subsection{Numerical results for the system with dissipation}
In this section, we evaluate how dissipation influences the dynamics of the kink in situations where the function $\lambda$  exhibits an explicit dependence on time. Such time-dependent variations in $\lambda$ significantly alter both qualitative and quantitative features of the kink’s motion, making their analysis essential for a complete understanding of the system’s behavior. In order to capture these effects, we focus on scenarios in which dissipative forces interplay with the temporal modulation of $\lambda$, potentially leading to nontrivial changes in the kink’s trajectory, velocity, and internal structure.
A central objective of this study is to verify the accuracy and applicability of the reduced description based on the effective model with two collective coordinates. This model, governed by the system of equations \eqref{2dof_ansatz_dis}, provides a simplified yet powerful framework for approximating the kink’s dynamics. By comparing its predictions against the expected physical behavior, we will assess its capability to reproduce the essential effects of dissipation in the presence of time-dependent driving. 

Figure \ref{nfig_15} shows the behavior of a kink that initially 
is positioned (with vanishing speed)
at the point $x_0=-12$. The dissipation coefficient is assumed to be equal to $\eta = 0.1$. The two main parameters describing the heterogeneity of the system are $\varepsilon_1 = 0.1$ and $\varepsilon_2 = 0.05$, respectively. Additionally, the function $\lambda$ takes the form of a traveling wave with a wavenumber $k=\frac{\pi}{6}$ and a frequency $\omega = 0.05$.
Panel (a) compares the trajectory obtained based on the field \eqref{phi4+} (solid black line) and that obtained from model 1 of Eq.~\eqref{2dof_ansatz_dis} (red dashed line). 
It can be seen that, as a result of competition between damping in the system, the presence of the barrier and the 
motion-inducing 
effect of the temporal and spatial dependence of the function $\lambda$,   the vibrations of the kink position do not occur around the minimum of the function $\mathcal{F}$, but around a new equilibrium position. The phenomena mentioned above prevent the kink from returning to its initial position. Note that the agreement between the two models is excellent even for times of the order of $t=500$. Using the same color codes, panel (b) compares the field model and model 1 for the variable $\gamma$ describing the inverse of the kink width. The agreement between the two models is very good, despite the complexity of the evolution
dynamics of the relevant variable. The graph shows that there are times when $\gamma$ falls below one. This effect has no kinematic (i.e., Lorentz-factor) basis (since the model is no longer Lorentz invariant) and is related to the broadening of the kink resulting from the reduction of the coupling constant.
\begin{figure}[h!]
    \centering
    \subfloat{{\includegraphics[height=4cm]{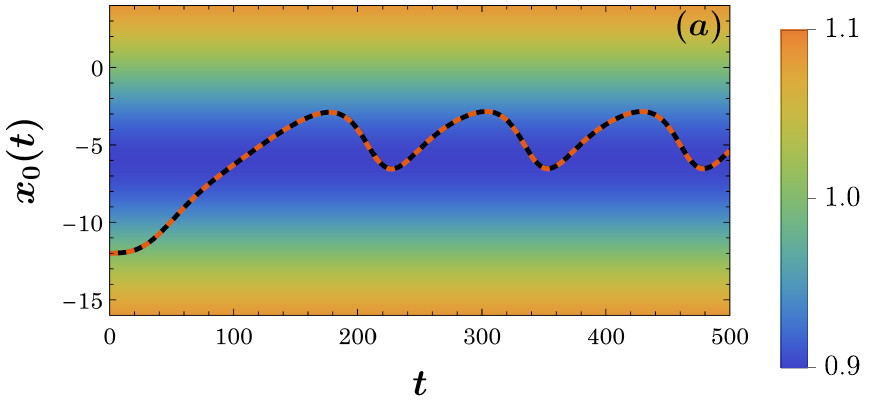}}}
    \quad
    \subfloat{{\includegraphics[height=4cm]{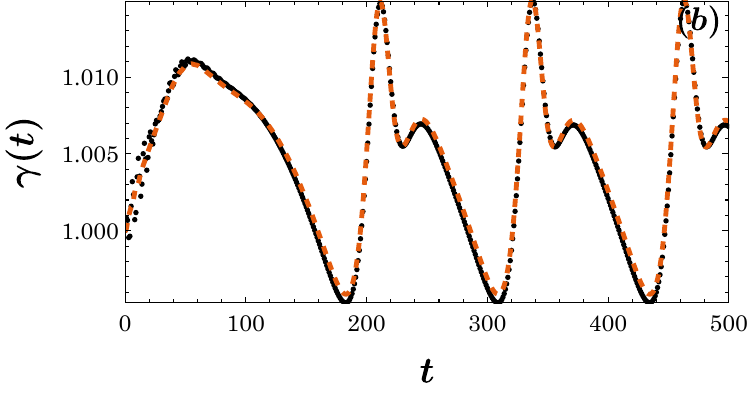}}}
    \caption{{Oscillations around a new equilibrium position in the presence of dissipation in the system. The following parameters are assumed in the figure: $\varepsilon_1 = 0.1$, $\varepsilon_2 = 0.05$, $k=\frac{\pi}{6}$, $\omega = 0.05$, $v=0$, $\eta = 0.1$, and $x_0=-12$. (a) The trajectory obtained from the field model \eqref{phi4+} is marked with a solid black line and compared with the trajectory obtained from the effective model \eqref{2dof_ansatz_dis} and marked with a red dashed line. 
    (b) Comparison of the time course of the variable $\gamma$ obtained from the field model  (black dotted line) with the course obtained from the approximate model (red dotted line).}}
    \label{nfig_15}
\end{figure}

In the next figure, i.e., Figure \ref{nfig_16} panel (a), in addition to the dissipation described by the coefficient $\eta = 0.1$, we have a temporarily activated forcing $\Gamma = 0.005$ in a form which, in a Josephson junction, is interpreted as a bias current. This excitation occurs from time $t=0$ to $t=150$ and is then switched off. The time interval where the force acts is marked in the figure by a slightly darkened (shaded) area.
 As a result of this forcing, the kink is pushed from the minimum at $x_0=-6$ (where it initially rests, i.e. $v=0$) to the neighboring minimum. There, it undergoes oscillations driven by the time dependence of the function $\lambda$, which pushes the kink toward the next barrier, while the dissipation present in the system restricts this motion so that the barrier is never crossed. In the figure, the function $\lambda$ is characterized by the following parameters: the amplitude of the changes is $\varepsilon_2=0.05$, the wavenumber is equal to $k=\frac{\pi}{6}$ while the frequency takes the value $\omega=0.05$. The function $\mathcal{F}$, which is associated with the spatial heterogeneity of the system, is characterized by the parameter $\varepsilon_1=0.1$. As the figure shows, the black solid line representing the kink trajectory obtained from the field solution is in excellent agreement with the red dashed line obtained from the effective model. The color bar on the right side of the figure refers to the values of the function $\mathcal{F}$. Panel (b) of this figure compares the evolution of variable $\gamma$ obtained from the field model {\color{red} \eqref{phi4+}} (black dotted line) and the effective model (red dotted line). The agreement between the two curves is again rather remarkable, especially given the rapid nature of the
 variation of the relevant quantity (especially near the turning points
 of its dynamical evolution). The slightly shaded area represents the time interval during which the external forcing $\Gamma$ is non-zero. The figure also shows that changes in the value of $\gamma$ are related not only to kinematic processes occurring in the system and associated with changes in velocity, but also (most notably in places where the value of $\gamma$ falls below one) to a temporary and local variation of the model coefficients, 
 causing the kink to broaden.
\begin{figure}[h!]
    \centering
    \subfloat{{\includegraphics[height=4cm]{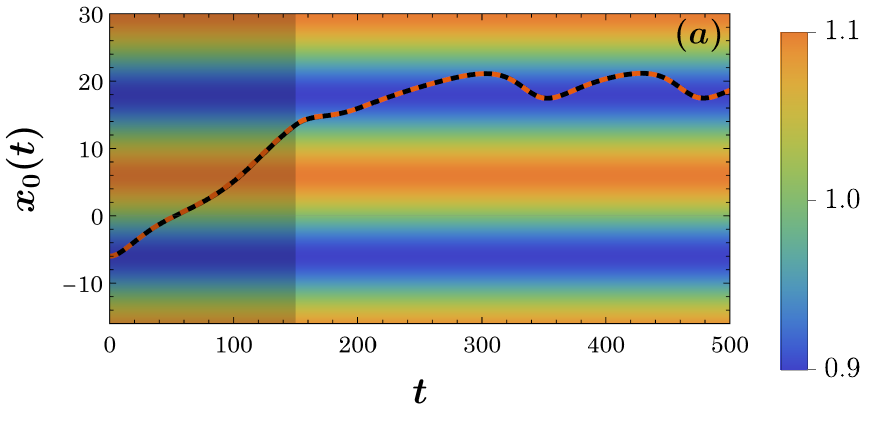}}}
    \quad
    \subfloat{{\includegraphics[height=4cm]{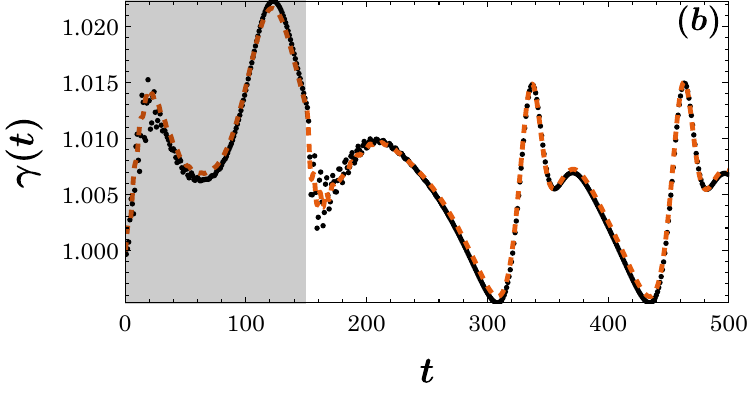}}}
    \caption{{External current as a control parameter that enables the transport of kink from one minimum to the adjacent minimum of the function $\mathcal{F}$. In the figure, the following parameters are assumed: $\varepsilon_1 = 0.1$, $\varepsilon_2 = 0.05$, $k=\frac{\pi}{6}$, $\omega = 0.05$, $v=0$, $\eta = 0.1$, $\Gamma = 0.005$ for $t \in [0,150]$ and $x_0=-6$. (a) The black solid line represents the result of the field model \eqref{phi4+}, while the red dashed line represents the trajectory obtained from the effective model \eqref{2dof_ansatz_dis}.  As usual, the color bar on the right refers to the values of the function $\mathcal{F}$. (b) Comparison of the time evolution of the variable $\gamma$ obtained based on the field model (black dotted line) and the trajectory obtained from the approximate model (red dotted line).} }
    \label{nfig_16}
\end{figure}

Figure \ref{nfig_17} illustrates in detail how the external forcing term $\Gamma$ allows to control the position of the kink within the network of minima defined by the function $\mathcal{F}$. The forcing is applied in two separate time windows:  $[ 100,250]$ and $[650,800]$.   In both panels of the figure, these time intervals are indicated by subtly shaded background regions, making it easier to visually correlate the periods of external action with the corresponding changes in the kink’s position. This visual coding highlights the direct link between the activation of the forcing term and the resulting shifts of the kink within the underlying landscape of the $\mathcal{F}$ function. Panel (a) of this figure shows the kink trajectory obtained from the field solution (continuous black line) and model 1 (red dashed line).  Despite the complex trajectory, the agreement between the two curves remains excellent. Note that initially, the external forcing in the form of the current $\Gamma$ shifts the kink to the neighboring minimum, where the kink begins to oscillate steadily. The impact of the second variation of $\Gamma$ is to displace the kink to the adjacent minimum, where it once again settles into stable oscillations. 
Panel (b) compares the time evolution $\gamma (t)$ for the field solution (black dotted line) and model 1 (red dotted line). Again, the agreement between the two plots is striking, even for times reaching $t=1000$. As before, the variable $\gamma$ falls below one, which is associated with a local decrease in the coupling constant $\lambda$, and is associated with the local temporal
broadening of the kink. It does not escape us that the above protocol provides a notion of a ``controller'' which allows for an efficient driving
of the kink trajectory, both at the level of the field model \eqref{phi4+} and ---accurately so, also--- at the level of the effective
non-conservative collective coordinate approach of Model 1 proposed as
the canonical approach to characterize the effective kink dynamics herein.
In applications where such ``targeted energy transfer'' (which has often
been argued as relevant in related nonlinear systems~\cite{aubrytargeted}) may be of interest, the
availability of such efficient techniques and associated 
effective descriptions may prove to be of
particular interest in the future.

\begin{figure}[h!]
    \centering
    \subfloat{{\includegraphics[height=4cm]{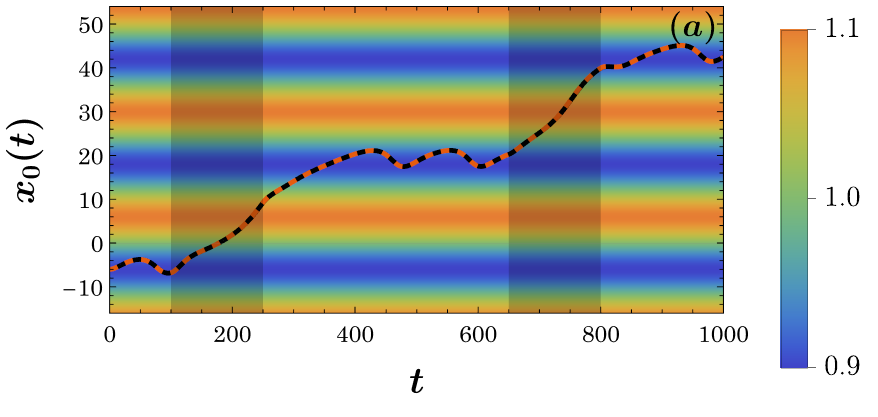}}}
    \quad
    \subfloat{{\includegraphics[height=4cm]{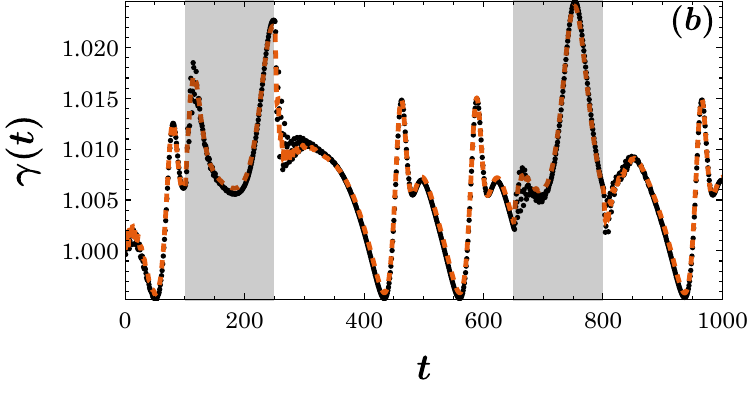}}}
    \caption{{Two successive transitions between the minima of the function $\mathcal{F}$, induced by an external forcing represented by a nonzero quantity $\Gamma$. Figure (a) compares the trajectory obtained from the field model \eqref{phi4+}, represented by a continuous black line, with the trajectory obtained from the approximate model \eqref{2dof_ansatz_dis}, marked by a red dashed line. As usual, the color bar on the right side of the figure refers to the values of the function $\mathcal{F}$. Panel (b) compares the $\gamma$ value obtained from the field model (black dotted line) and the approximate model (red dotted line). The gray areas on both panels show the time during which a non-zero forcing occurs. The parameters in the figure are as follows: $\varepsilon_1 = 0.1$, $\varepsilon_2 = 0.05$, $k=\frac{\pi}{6}$, $\omega = 0.05$, $v=0$, $\eta = 0.1$, $\Gamma = 0.005$ for $t \in [100,250] \cup [650,800]$ and $x_0=-6$.} }
    \label{nfig_17}
\end{figure}

\section{Conclusions and Future Challenges}
In the present work, we have analyzed the dynamics of a kink in the $\phi^4$ model with spatial inhomogeneities, explicitly breaking translational invariance
(and time-reversal symmetry). Particular attention was devoted to the effects of spatial variations of the model coefficients, and furthermore the role
of dissipation and drive was also considered. The explicit time dependence of the coefficients makes the system non-autonomous, which substantially complicates the kink dynamics. We used Fourier modes for the variation of 
model coefficients, bearing in mind that more complex such
variations can be decomposed into such Fourier mode building blocks.

The main objective was to construct a reliable effective description that reproduces the kink behavior faithfully over long evolution times. The method proposed, previously developed in the context of the sine-Gordon model \cite{Dobrowolski2025}, was extended to account for spatial inhomogeneities. We introduced three effective descriptions of the kink: model 1 based on the kink position and inverse width $(x_0,\gamma)$ and models 2 and 3 based on $(x_0,b)$, the kink position and amplitude of the discrete internal mode.

In the autonomous case, all three reduced models showed very good agreement with the solution of the field model, even for time intervals of several hundred units. When the coupling constant is allowed to vary in time, however, the situation changes markedly. Models 2 and 3 that rely on the dynamical variables $(x_0,b)$ capture the kink dynamics only within a narrow temporal window, while model 1 remains accurate over a broader temporal (and frequency) range. Nonetheless, the accuracy of model 1 also deteriorates at high modulation frequencies of the coupling constant $\lambda$. One can attribute this to the suitable scaling 
of the width (and incorporation of the position) of the kink in model 1.
The relevant nonlinear waveform in model 1 has been argued to be more
suitable than the linearization-inspired ans{\"a}tze of models 2 and 3.

Let us note that both the present work and the articles \cite{Dobrowolski2025, Dobrowolski2025b} point toward a clear path for constructing reliable and highly accurate effective models capable of describing soliton dynamics. These models reduce the description of soliton motion to only a few degrees of freedom, which capture even the complex dynamics of field systems with explicit breaking of translational (and time-reversal) symmetry. The proposed method also provides a framework for treating non-autonomous systems subject to dissipation
and external drive. It is anticipated that investigations along these lines will be of increased relevance in the future, including for purposes associated with
control and targeted transfer of the relevant nonlinear kink excitations.
In the present setting we demonstrated some prototypical paradigms of
such control.

It is worth noting that the issue of descriptive accuracy has already been of particular relevance to studies of kink–antikink interactions. These investigations have demonstrated that the standard ansatz based on a discrete mode can be freed from the so-called null-vector 
problem~\cite{Manton2021}; however, the quantitative reconstruction of the resonance structure remains an open challenge. To a large extent, an improvement in the quantitative description has been achieved through an ansatz incorporating Derrick modes, as proposed in~\cite{Adam2022}. In our case, the dynamics involves multiple effects associated with the presence of inhomogeneities in the system, as well as its non-autonomous character but for the case of a single kink. Our results suggest that a significantly more accurate description of the resulting
dynamics can be obtained within a framework of ansätze naturally linked to Derrick modes.

A further natural avenue of investigation is the inclusion of additional spatial dimensions. The kink, as a solution interpolating between two vacuum states, can be interpreted as a domain wall in three spatial dimensions or as a string separating regions of different vacua in two spatial dimensions. In higher-dimensional settings with inhomogeneities, it becomes particularly interesting to investigate the dynamics of scalar field zero surfaces (in three dimensions) or scalar field zero lines (in two dimensions) and explore their reduced-dimension dynamics. 
In each $d$-dimensional case, the relevant goal is to obtain a reduced $d-1$-dimensional
effective description.
Studies of this kind have been carried out, for example, in the context of the sine-Gordon model in \cite{Gatlik2024}. It should be noted, however, that these investigations were based on the standard ansatz rather than on one that explicitly accounts for spatial inhomogeneity of the system. Developing a systematic
approach for spatially and temporally varying coefficients, including
in the presence of damping and driving would be an interesting direction
for future studies. 

\FloatBarrier
\section*{Acknowledgments}
We gratefully acknowledge Polish high-performance computing infrastructure PLGrid (HPC Center: ACK Cyfronet AGH) for providing computer facilities and support within computational grant no. PLG/2025/018503 (JG).
This material is partially based upon work supported by the U.S. National Science Foundation under the award PHY-2408988 (PGK). This research was partly conducted while P.G.K. was 
visiting the Okinawa Institute of Science and
Technology (OIST) through the Theoretical Sciences Visiting Program (TSVP)
and the Sydney Mathematical Research Institute (SMRI) under an SMRI Visiting
Fellowship. 
This work was also supported by the Simons Foundation
[SFI-MPS-SFM-00011048, P.G.K.].

\appendix
\renewcommand{\theequation}{\thesection\arabic{equation}}
\setcounter{equation}{0}

\section{Appendix: Coefficients of model 1}
\label{AppendixA}
The coefficients featured in the Lagrangian expression \eqref{Leff} are not constants; rather, they vary as explicit functions of the collective variables $x_0$, $\gamma$, as well as the temporal parameter $t$. These dependencies are specified through a set of integral formulas presented below, which encapsulate how each coefficient evolves in response to changes in the system’s configuration and its progression in time. Such integral definitions are essential for capturing the interplay between spatial structure and dynamic variables within the reduced description framework
\begin{equation}
\begin{gathered}
\label{integrals}
    M = \int_{-\infty}^{+\infty}dx ~\sech^4(\xi) \, W^2(\xi),\\
    m = \frac{1}{\gamma^2}\int_{-\infty}^{+\infty}dx ~ \sech^4(\xi) \, \xi^2,\\
    \kappa = \frac{1}{\gamma} \int_{-\infty}^{+\infty}dx~\sech^4(\xi) W(\xi) \, \xi ,\\
\alpha = \int_{-\infty}^{+\infty}dx ~  W(\xi) \, \xi \,\sech^4(\xi) \,  \frac{\partial_t \lambda(t,x_0)}{2 \lambda(t,x_0)}  
    , \\
\beta = \frac{1}{\gamma} \int_{-\infty}^{+\infty}dx ~ \xi^2 \sech^4(\xi) \,   \frac{\partial_t \lambda(t,x_0)}{2 \lambda(t,x_0)}   
    ,\\
    V = \int_{-\infty}^{+\infty}dx \,  \sech^4(\xi) \,  \left(  \frac{1}{2} \, \lambda(t,x_0) \gamma^2 \,\frac{ \mathcal{F}(x)}{ 2 \mathcal{F}(x_0)} - \frac{1}{2} \, \xi^2 \, \left(\frac{ \partial_t \lambda(t,x_0)}{ 2 \lambda(t,x_0)}\right)^2 + \frac{1}{4} \lambda(t,x) \right)  .
\end{gathered}
\end{equation}
The auxiliary function $W$ in the above expressions has the form 
\begin{equation} \label{W}
   W(\xi) = \xi N - \gamma \sqrt{\frac{\lambda(t,x_0)}{2 {\cal F}(x_0)}}
    , \,\,\,{\mathrm{where}}\,\,\,
    N = \frac{1}{2} \, \left(   \frac{\partial_{x_0} { \lambda}(t,x_0)}{ {\lambda}(t,x_0)} - \frac{\partial_{x_0} {\cal F}(x_0)}{ {\cal F}(x_0)} \right)     .
\end{equation}
The majority of the integrals presented above admit explicit analytical evaluation, allowing for direct and precise expressions to be obtained. However, there is an exception in the case of the potential term, which involves spatially dependent functions such as $\mathcal{F}(x)$ and the time- and space-dependent coefficient $\lambda(t,x)$. By retaining the terms involving these functions in integral form, the formulation remains applicable to arbitrary inhomogeneities characterized by $\lambda$ and ${\mathcal F}$
\begin{equation}
\begin{gathered}
\label{integrals+c}
    M =\frac{2}{ 3 \gamma}\, \sqrt{\frac{2 \mathcal{F}(x_0)}{\lambda(t,x_0)}} \left[ \left( \frac{\pi^2}{6} - 1 \right) N^2 + \gamma^2 \,\frac{\lambda(t,x_0)}{ \mathcal{F}(x_0)}\right],\\
    m = \frac{1}{  \gamma}\, \sqrt{\frac{2 \mathcal{F}(x_0)}{\lambda(t,x_0)}} \, \left(\frac{\pi^2}{9} - \frac{2}{3} \right) \,  ,\\
\kappa = \frac{1}{  \gamma^2}\, \sqrt{\frac{2 \mathcal{F}(x_0)}{\lambda(t,x_0)}} \, \left(\frac{\pi^2}{9} - \frac{2}{3} \right) N
    , \\
\alpha = \frac{1}{  \gamma}\, \sqrt{\frac{2 \mathcal{F}(x_0)}{\lambda(t,x_0)}} \, \left(\frac{\pi^2}{9} - \frac{2}{3} \right)  N  \,\, \frac{\partial_t \lambda(t,x_0)}{2 \lambda(t,x_0)}
    , \\
\beta = \frac{1}{  \gamma^2}\, \sqrt{\frac{2 \mathcal{F}(x_0)}{\lambda(t,x_0)}} \, \left(\frac{\pi^2}{9} - \frac{2}{3} \right)    \, \frac{\partial_t \lambda(t,x_0)}{2 \lambda(t,x_0)}  
    ,\\
    V = \frac{1}{4} \int_{-\infty}^{+\infty}dx \,  \sech^4(\xi) \,  \left(\lambda(t,x) +  \lambda(t,x_0) \,\gamma^2 \,\frac{ \mathcal{F}(x)}{ \mathcal{F}(x_0)}     \right) -  \frac{1}{  3 \gamma}\, \sqrt{\frac{2 \mathcal{F}(x_0)}{\lambda(t,x_0)}} \, \left(\frac{\pi^2}{6} - 1 \right)    \, \left( \frac{\partial_t \lambda(t,x_0)}{2 \lambda(t,x_0)} \right)^2 .
\end{gathered}
\end{equation}

\setcounter{equation}{0}
\section{Appendix: Coefficients of model 3}
\label{AppendixB}
All coefficients present in the Lagrangian \eqref{Leff+} are functions of the $x_0$, $b$, and $t$ variables. They  are defined by the integrals below
\begin{equation}
\begin{gathered}
\label{integrals+}
    M_b = \int_{-\infty}^{+\infty}dx ~{\mathcal{K}^2} \, W^2(\xi),\\
    m_b = \int_{-\infty}^{+\infty}dx ~ \tanh^2(\xi) \sech^2(\xi) ,\\
    \kappa_b = \int_{-\infty}^{+\infty}dx~\mathcal{K} W(\xi) \tanh(\xi) \sech(\xi)   ,\\
\alpha_b = \int_{-\infty}^{+\infty}dx ~ \mathcal{K}^2 W(\xi) \, \xi \, \,  \frac{\partial_t \lambda(t,x_0)}{2 \lambda(t,x_0)}  
    , \\
\beta_b =  \int_{-\infty}^{+\infty}dx ~ \mathcal{K} \xi \, \tanh(\xi) \sech(\xi)  \frac{\partial_t \lambda(t,x_0)}{2 \lambda(t,x_0)}   
    ,\\
    V_b = \int_{-\infty}^{+\infty}dx \,   \,  \left[  \frac{1}{2} \, \lambda(t,x_0) \mathcal{K}^2 \,\frac{ \mathcal{F}(x)}{ 2 \mathcal{F}(x_0)} - \frac{1}{2} \, \mathcal{K}^2 \, \xi^2 \, \left(\frac{ \partial_t \lambda(t,x_0)}{ 2 \lambda(t,x_0)}\right)^2 + \frac{1}{4} \lambda(t,x) \left( \tanh^2(\xi) \left( 1 + b \sech(\xi)\right)^2-1 \right)^2\right]  .
\end{gathered}
\end{equation}
The auxiliary functions which appear within the expressions discussed above are explicitly constructed in the following manner
\begin{equation} \label{W+}
   W(\xi) = \xi N -  \sqrt{\frac{\lambda(t,x_0)}{2 {\cal F}(x_0)}}
    , \,\,\,{\mathrm{where}}\,\,\,
    N = \frac{1}{2} \, \left(   \frac{\partial_{x_0} { \lambda}(t,x_0)}{ {\lambda}(t,x_0)} - \frac{\partial_{x_0} {\cal F}(x_0)}{ {\cal F}(x_0)} \right)     ,
\end{equation}
and
\begin{equation}
    \mathcal{K} = \sech^2 (\xi) + b \sech(\xi) \left( -1 + 2 \sech^2 (\xi) \right) .
\end{equation}
In the effective potential, some terms have been integrated explicitly, while those depending on the functions of the spatial variable i.e. ${\mathcal F}(x)$ and $\lambda(t,x)$ have been left in integral form
\begin{equation}
\begin{gathered}
\label{integrals+c+}
    M_b = \sqrt{\frac{2 {\mathcal{F}}}{\lambda}} \left\{ \frac{\lambda}{\mathcal{F}} \left( \frac{2}{3}+\frac{7}{15} b^2 + \frac{\pi}{4} b \right) +\mathcal{N}^2 \left[-\frac{2}{3} + \frac{\pi^2}{9} + \pi b \left(-\frac{4}{3} + \frac{7}{90} \pi b+\frac{\pi^2}{8} \right) \right] \right\} 
    ,\\
    m_b = \frac{2}{3} \, \sqrt{\frac{2 {\mathcal{F}}}{\lambda}} ,\\
    \kappa_b = \frac{1}{3} \,\sqrt{\frac{2 {\mathcal{F}}}{\lambda}} \mathcal{N} \left(\frac{\pi}{2} -b\right)  ,\\
\alpha_b =   \sqrt{\frac{2 {\mathcal{F}}}{\lambda}} \mathcal{N} \,\frac{\partial_t \lambda(t,x_0)}{2 \lambda(t,x_0)} \left[ \left( \frac{\pi^2}{9} - \frac{2}{3} \right) + \frac{7 \pi^2}{90} b^2 + \left( \frac{\pi^3}{8} - \frac{4 \pi}{3} \right) b\right],
    \\
\beta_b = \frac{1}{3}\, \sqrt{\frac{2 {\mathcal{F}}}{\lambda}} \, \frac{\partial_t \lambda(t,x_0)}{2 \lambda(t,x_0)} \left( \frac{\pi}{2} - b \right)  
    ,\\
    V_b = \frac{1}{4}\int_{-\infty}^{+\infty}dx \,   \,  \left[  \, \lambda(t,x_0) \mathcal{K}^2 \,\frac{ \mathcal{F}(x)}{ \mathcal{F}(x_0)}  + \lambda(t,x) \left( \tanh^2(\xi) \left( 1 + b \sech(\xi)\right)^2-1 \right)^2\right]  
   + \\ \frac{1}{6} \,\sqrt{\frac{2 {\mathcal{F}}}{\lambda}}  \left(\frac{ \partial_t \lambda(t,x_0)}{ 2 \lambda(t,x_0)}\right)^2 \left[2 - \frac{\pi^2}{3} + \pi b \left(4 - \frac{7}{30} \pi b - \frac{3}{8} \pi^2 \right) \right].
\end{gathered}
\end{equation}
\FloatBarrier

\setcounter{equation}{0}
\FloatBarrier
\section{Appendix: Comparison of Derrick's mode present in model 1 with the vibration modes of model 3 and 2. }
\label{AppendixC}
In this appendix, we will compare Derrick's mode, which is the first term of the expansion of ansatz 1, defined by the equations \eqref{kink-xi} and \eqref{xi} in the gamma variable around the value of unity, with the vibration modes that are part of ansatz \eqref{kink-xi2++}, \eqref{xi2++} (which forms the basis of model 2) and ansatz \eqref{kink-xi2}, \eqref{xi2+} (which defines the model 3).
Expanding the ansatz defining model 1 in variable $\gamma$, around the value one, we obtain:
\begin{equation}
\begin{gathered}\label{xi=}
 \phi(t,x) = \tanh \left( \sqrt{ \frac{\lambda(t,x_0)}{2 {\cal
F}(x_0)}} \,\, \gamma \, (x-x_0) \right) \approx \tanh \left( \sqrt{ \frac{\lambda(t,x_0)}{2 {\cal
F}(x_0}} \,\,  \, (x-x_0) \right)+ \\ \sqrt{ \frac{\lambda(t,x_0)}{2 {\cal
F}(x_0)}} \,\,  \, (x-x_0) \sech^2 \left( \sqrt{ \frac{\lambda(t,x_0)}{2 {\cal
F}(x_0)}} \,\,  \, (x-x_0) \right) (\gamma-1) .
\end{gathered}
\end{equation}
From this expansion, we can read Derrick's mode
\begin{equation}
\label{Derrick}
\psi_1(t,x) = (\gamma-1) \sqrt{ \frac{\lambda(t,x_0)}{2 {\cal
F}(x_0)}} \,\,  \, (x-x_0) \sech^2 \left( \sqrt{ \frac{\lambda(t,x_0)}{2 {\cal
F}(x_0)}} \,\,  \, (x-x_0) .\right)   .
\end{equation}
On the other hand, the vibration mode constituting the second term of the ansatz defining model 3 has the form:
\begin{equation}
\label{vib3}
\psi_3(t,x) =   b \tanh \left( \sqrt{ \frac{\lambda(t,x_0)}{2 {\cal
F}(x_0)}} \,\,  \, (x-x_0) \right)  \sech \left( \sqrt{ \frac{\lambda(t,x_0)}{2 {\cal
F}(x_0)}} \,\,  \, (x-x_0) \right)    .
\end{equation}
Similarly, for the second model, the vibration mode takes the form:
\begin{equation}
\label{vib2}
\psi_2(t,x) =   b \tanh \left( \sqrt{ \frac{1}{2} }\,\,  \, (x-x_0) \right)  \sech \left( \sqrt{ \frac{1}{2}} \,\,  \, (x-x_0) \right)    .
\end{equation}
To highlight the similarities and differences between these three models, Figure \ref{fig_App01} shows graphs representing the above functions. In the figure, the solid red line shows the shape of Derrick's mode. The dashed lines depict the vibration modes of model 3 (green) and model 2 (purple). The basis for this comparison is, on the one hand, figures \ref{nfig_05} and  \ref{nfig_06} in panel (a), where the parameter values and initial conditions shown in these figures were adopted. On the other hand, by assuming the parameter values and initial conditions from figures \ref{nfig_07} and \ref{nfig_08}, we obtain the comparison presented in panel (b). Since in formulas \eqref{Derrick}, \eqref{vib3} and \eqref{vib2} $x_0$, $b$, and $\gamma$ vary over time, we used the time-averaged values of these functions for the comparison. Panel (a) presents the results obtained after averaging (in 
order to provide a tangible yardstick for comparison), with the mean values given by $\bar{x}_0 = -4.71$, $\bar{\gamma} = 1.07$, $\bar{b} = 0.04$ for model 3, and $\bar{b} = -0.05$ for model 2. It can be noticed that the vibration mode of model 2, in this case, fits better into the shape of Derrick's mode than the vibration mode of model 3. This result corresponds to the trajectory in figures \ref{nfig_05} and  \ref{nfig_06}, where the trajectory obtained based on model 2 corresponds better to the trajectory of the field model. For comparison, panel (b) shows the results corresponding to the parameter values used in figures \ref{nfig_07} and  \ref{nfig_08}, where the calculated mean values of the dynamic variables are $\bar{x}_0 = -5.98$, $\bar{\gamma} = 1.08$, $\bar{b} = 0.05$ for model 3, and $\bar{b} = 0.17$ for model 2. This time, the vibration mode of model 3 aligns more closely with Derrick's mode than that of model 2.
This behavior corresponds well with the results shown in Figures \ref{nfig_07}-\ref{nfig_08}, where the trajectory of model 3 better matches the field configuration than the trajectory of model 2.

\begin{figure}[h!]
    \centering
    \subfloat{{\includegraphics[height=4cm]{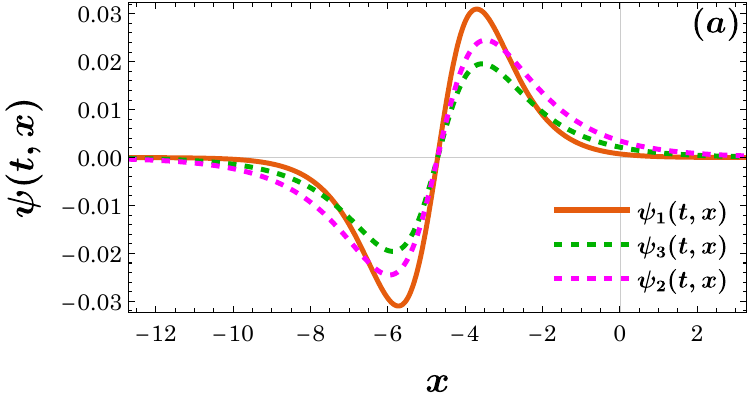}}}
    \quad
    \subfloat{{\includegraphics[height=4cm]{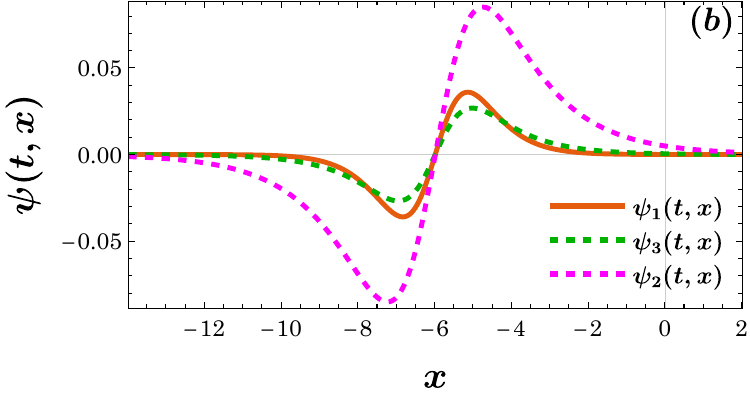}}}
    \caption{(a) Results obtained with the parameter values and initial conditions as in figures \ref{nfig_05} and \ref{nfig_06}. The time-averaged values of the dynamic variables are $\bar{x}_0 = -4.71$, $\bar{\gamma} = 1.07$, $\bar{b} = 0.04$ for model 3, and $\bar{b} = -0.05$ for model 2. (b) Results obtained with the parameter values and initial conditions as in figures \ref{nfig_07} and \ref{nfig_08}. The time-averaged values of the dynamic variables are $\bar{x}_0 = -5.98$, $\bar{\gamma} = 1.08$, $\bar{b} = 0.05$ for model 3, and $\bar{b} = 0.17$ for model 2.} 
    \label{fig_App01}
\end{figure}

\printbibliography

@PREAMBLE{
 "\providecommand{\noopsort}[1]{}" 
 # "\providecommand{\singleletter}[1]{#1}%" 
}

@book{p4book,
  title     = {A Dynamical Perspective on the $\phi^4$ Model},
  editor    = {Cuevas-Maraver, Jesús and Kevrekidis, Panayotis G.},
  publisher = {Springer International Publishing},
  series    = {Nonlinear Systems and Complexity},
  year      = {2019},
  edition   = {1st},
  isbn      = {978-3030118389}
}

@article{aubrytargeted,
  title = {Targeted Energy Transfer through Discrete Breathers in Nonlinear Systems},
  author = {Kopidakis, G. and Aubry, S. and Tsironis, G. P.},
  journal = {Phys. Rev. Lett.},
  volume = {87},
  issue = {16},
  pages = {165501},
  numpages = {4},
  year = {2001},
  month = {9},
  publisher = {American Physical Society},
  doi = {10.1103/PhysRevLett.87.165501},
  url = {https://link.aps.org/doi/10.1103/PhysRevLett.87.165501}
}

@article{Xin2000Fronts,
  author  = {Xin, Jack},
  title   = {Front Propagation in Heterogeneous Media},
  journal = {SIAM Review},
  volume  = {42},
  number  = {2},
  pages   = {161--230},
  year    = {2000},
  doi     = {10.1137/S0036144599364296},
  url     = {https://www.math.uci.edu/~jxin/xinsrev00.pdf}
}

@article{Galley2013,
  title = {Classical Mechanics of Nonconservative Systems},
  author = {Galley, Chad R.},
  journal = {Phys. Rev. Lett.},
  volume = {110},
  issue = {17},
  pages = {174301},
  numpages = {5},
  year = {2013},
  publisher = {American Physical Society},
  doi = {10.1103/PhysRevLett.110.174301},
  url = {https://link.aps.org/doi/10.1103/PhysRevLett.110.174301}
}

@article{Kevrekidis2014,
  title = {Variational method for nonconservative field theories: Formulation and two $\mathcal{PT}$-symmetric case examples},
  author = {Kevrekidis, P. G.},
  journal = {Phys. Rev. A},
  volume = {89},
  issue = {1},
  pages = {010102},
  numpages = {5},
  year = {2014},
  publisher = {American Physical Society},
  doi = {10.1103/PhysRevA.89.010102},
  url = {https://link.aps.org/doi/10.1103/PhysRevA.89.010102}
}

@article{Dobrowolski2025,
  title = {Kink movement on a periodic background},
  author = {Dobrowolski, Tomasz and Gatlik, Jacek and Kevrekidis, Panayotis G.},
  journal = {Phys. Rev. E},
  volume = {111},
  issue = {2},
  pages = {024203},
  numpages = {18},
  year = {2025},
  month = {2},
  publisher = {American Physical Society},
  doi = {10.1103/PhysRevE.111.024203},
  url = {https://link.aps.org/doi/10.1103/PhysRevE.111.024203}
}

@inbook{Chaikin1995, 
place={Cambridge}, 
title={Walls, kinks and solitons}, 
booktitle={Principles of Condensed Matter Physics},
publisher={Cambridge University Press},
author={Chaikin, P. M. and Lubensky, T. C.},
year={1995}, 
pages={590–661}}

@book{Vachaspati2023, 
place={Cambridge}, 
title={Kinks and Domain Walls: An Introduction to Classical and Quantum Solitons}, 
publisher={Cambridge University Press}, 
author={Vachaspati, Tanmay}, 
year={2023}}

@book{Rajaraman1982,
  author    = {R. Rajaraman},
  title     = {Solitons and Instantons: An Introduction to Solitons and Instantons in Quantum Field Theory},
  publisher = {North-Holland Publishing Company},
  address   = {Amsterdam},
  year      = {1982}
}

@article{Catalan2012,
  title = {Domain wall nanoelectronics},
  author = {Catalan, G. and Seidel, J. and Ramesh, R. and Scott, J. F.},
  journal = {Rev. Mod. Phys.},
  volume = {84},
  issue = {1},
  pages = {119--156},
  numpages = {0},
  year = {2012},
  month = {2},
  publisher = {American Physical Society},
  doi = {10.1103/RevModPhys.84.119},
  url = {https://link.aps.org/doi/10.1103/RevModPhys.84.119}
}

@article{Meier2021,
    author = {Meier, Dennis and Valanoor, Nagarajan and Zhang, Qi and Lee, Donghwa},
    title = {Domains and domain walls in ferroic materials},
    journal = {Journal of Applied Physics},
    volume = {129},
    number = {23},
    pages = {230401},
    year = {2021},
    month = {06},
    issn = {0021-8979},
    doi = {10.1063/5.0057144},
    url = {https://doi.org/10.1063/5.0057144},
    eprint = {https://pubs.aip.org/aip/jap/article-pdf/doi/10.1063/5.0057144/15262276/230401\_1\_online.pdf},
}

@article{Salje2012,
   author = "Salje, Ekhard K.H.",
   title = "Ferroelastic Materials", 
   journal= "Annual Review of Materials Research",
   year = "2012",
   volume = "42",
   number = "Volume 42, 2012",
   pages = "265-283",
   doi = "https://doi.org/10.1146/annurev-matsci-070511-155022",
   url = "https://www.annualreviews.org/content/journals/10.1146/annurev-matsci-070511-155022",
   publisher = "Annual Reviews",
   issn = "1545-4118",
   type = "Journal Article",
  }

@Article{Bertoldi2017,
author={Bertoldi, Katia
and Vitelli, Vincenzo
and Christensen, Johan
and van Hecke, Martin},
title={Flexible mechanical metamaterials},
journal={Nature Reviews Materials},
year={2017},
month={10},
day={17},
volume={2},
number={11},
pages={17066},
issn={2058-8437},
doi={10.1038/natrevmats.2017.66},
url={https://doi.org/10.1038/natrevmats.2017.66}
}

@book{Seidel2016,
  editor       = {Jan Seidel},
  title        = {Topological Structures in Ferroic Materials},
  subtitle     = {Domain Walls, Vortices and Skyrmions},
  series       = {Springer Series in Materials Science},
  volume       = {228},
  publisher    = {Springer},
  address      = {Cham},
  year         = {2016},
  doi          = {10.1007/978-3-319-25301-5},
  isbn         = {978-3-319-25299-5},
  issn         = {0933-033X},
  eissn        = {2196-2812},
  edition      = {1},
  pages        = {XII+241},
  note         = {46 b/w illustrations, 75 illustrations in colour}
}

@article{Nataf2020,
doi = {10.1088/1361-648X/ab68f3},
url = {https://dx.doi.org/10.1088/1361-648X/ab68f3},
year = {2020},
month = {2},
publisher = {IOP Publishing},
volume = {32},
number = {18},
pages = {183001},
author = {Nataf, G F and Guennou, M},
title = {Optical studies of ferroelectric and ferroelastic domain walls},
journal = {Journal of Physics: Condensed Matter}
}

@article{Saarloos2003,
title = {Front propagation into unstable states},
journal = {Physics Reports},
volume = {386},
number = {2},
pages = {29-222},
year = {2003},
issn = {0370-1573},
doi = {https://doi.org/10.1016/j.physrep.2003.08.001},
url = {https://www.sciencedirect.com/science/article/pii/S0370157303003223},
author = {Wim {van Saarloos}}
}

@article{Hahn2005,
  title = {The mean-field ${\ensuremath{\varphi}}^{4}$ model: Entropy, analyticity, and configuration space topology},
  author = {Hahn, Ingo and Kastner, Michael},
  journal = {Phys. Rev. E},
  volume = {72},
  issue = {5},
  pages = {056134},
  numpages = {9},
  year = {2005},
  month = {11},
  publisher = {American Physical Society},
  doi = {10.1103/PhysRevE.72.056134},
  url = {https://link.aps.org/doi/10.1103/PhysRevE.72.056134}
}

@article{Gatlik2024,
  title = {Effective description of the impact of inhomogeneities on the movement of the kink front in $2+1$ dimensions},
  author = {Gatlik, Jacek and Dobrowolski, Tomasz and Kevrekidis, Panayotis G.},
  journal = {Phys. Rev. E},
  volume = {109},
  issue = {2},
  pages = {024205},
  numpages = {20},
  year = {2024},
  month = {2},
  publisher = {American Physical Society},
  doi = {10.1103/PhysRevE.109.024205},
  url = {https://link.aps.org/doi/10.1103/PhysRevE.109.024205}
}

@article{Braghin1998,
  title = {Time dependent variational calculations for the quantum fluctuations of the $\ensuremath{\lambda}{\ensuremath{\varphi}}^{4}$ model},
  author = {Braghin, F\'abio L.},
  journal = {Phys. Rev. D},
  volume = {57},
  issue = {6},
  pages = {3548--3557},
  numpages = {0},
  year = {1998},
  month = {3},
  publisher = {American Physical Society},
  doi = {10.1103/PhysRevD.57.3548},
  url = {https://link.aps.org/doi/10.1103/PhysRevD.57.3548}
}

@article{Destri2000,
  title = {Improved time-dependent Hartree-Fock approach for scalar ${\ensuremath{\varphi}}^{4}$ QFT},
  author = {Destri, C. and Manfredini, E.},
  journal = {Phys. Rev. D},
  volume = {62},
  issue = {2},
  pages = {025008},
  numpages = {21},
  year = {2000},
  month = {6},
  publisher = {American Physical Society},
  doi = {10.1103/PhysRevD.62.025008},
  url = {https://link.aps.org/doi/10.1103/PhysRevD.62.025008}
}

@article{Dobrowolski2025b,
    author = {Dobrowolski, Tomasz and Gatlik, Jacek and Bryłowska, Zofia and Kevrekidis, Panayotis G.},
    title = {Kink dynamics in a non-autonomous sine-Gordon model},
    journal = {Chaos: An Interdisciplinary Journal of Nonlinear Science},
    volume = {35},
    number = {10},
    pages = {103128},
    year = {2025},
    month = {10},
    issn = {1054-1500},
    doi = {10.1063/5.0284611},
    url = {https://doi.org/10.1063/5.0284611},
    eprint = {https://pubs.aip.org/aip/cha/article-pdf/doi/10.1063/5.0284611/20761493/103128_1_5.0284611.pdf},
}

@article{Adam2022,
  title = {Relativistic moduli space for kink collisions},
  author = {Adam, C. and Manton, N. S. and Oles, K. and Romanczukiewicz, T. and Wereszczynski, A.},
  journal = {Phys. Rev. D},
  volume = {105},
  issue = {6},
  pages = {065012},
  numpages = {18},
  year = {2022},
  month = {3},
  publisher = {American Physical Society},
  doi = {10.1103/PhysRevD.105.065012},
  url = {https://link.aps.org/doi/10.1103/PhysRevD.105.065012}
}

@article{Manton2021,
  title = {Kink moduli spaces: Collective coordinates reconsidered},
  author = {Manton, N. S. and Ole\ifmmode \acute{s}\else \'{s}\fi{}, K. and Roma\ifmmode \acute{n}\else\'{n}\fi{}czukiewicz, T. and Wereszczy\ifmmode \acute{n}\else\'{n}\fi{}ski, A.},
  journal = {Phys. Rev. D},
  volume = {103},
  issue = {2},
  pages = {025024},
  numpages = {20},
  year = {2021},
  month = {1},
  publisher = {American Physical Society},
  doi = {10.1103/PhysRevD.103.025024},
  url = {https://link.aps.org/doi/10.1103/PhysRevD.103.025024}
}

@article{Ma2025,
  title = {Radial kinks in boson stars},
  author = {Ma, Tian-Chi and Wang, Xiang-Yu and Zhang, Hai-Qing},
  journal = {Phys. Rev. D},
  volume = {113},
  issue = {4},
  pages = {044066},
  numpages = {11},
  year = {2026},
  month = {Feb},
  publisher = {American Physical Society},
  doi = {10.1103/yw4x-ncmh},
  url = {https://link.aps.org/doi/10.1103/yw4x-ncmh}
}

@article{Caputo2024,
  title = {Radial kinks in a Schwarzschild-like geometry},
  author = {Caputo, Jean-Guy and Dobrowolski, Tomasz and Gatlik, Jacek and Kevrekidis, Panayotis G.},
  journal = {Phys. Rev. D},
  volume = {110},
  issue = {12},
  pages = {125025},
  numpages = {17},
  year = {2024},
  month = {12},
  publisher = {American Physical Society},
  doi = {10.1103/PhysRevD.110.125025},
  url = {https://link.aps.org/doi/10.1103/PhysRevD.110.125025}
}

@article{Takyi2016,
  title = {Collective coordinates in one-dimensional soliton models revisited},
  author = {Takyi, I. and Weigel, H.},
  journal = {Phys. Rev. D},
  volume = {94},
  issue = {8},
  pages = {085008},
  numpages = {11},
  year = {2016},
  month = {10},
  publisher = {American Physical Society},
  doi = {10.1103/PhysRevD.94.085008},
  url = {https://link.aps.org/doi/10.1103/PhysRevD.94.085008}
}

\end{document}